\documentclass[pre,aps,onecolumn,superscriptaddress,nofootinbib]{revtex4-1}
\pdfoutput=1
\usepackage{amsmath, amsthm, amssymb}
\usepackage{amsfonts}
\usepackage{graphicx}
\usepackage{dcolumn}
\usepackage{bm}
\usepackage{textcomp}
\usepackage[normalem]{ulem}

\usepackage[titletoc,title]{appendix}

\usepackage{color}

\usepackage{colordvi}
\usepackage[usenames,dvipsnames]{xcolor}

\usepackage{epsfig}
\usepackage{textcomp}
\usepackage{epstopdf}

\usepackage{amsmath, amsthm, amssymb}
\usepackage{bm}

\usepackage[colorlinks=true, urlcolor=blue, anchorcolor=blue, citecolor=blue,filecolor=blue,linkcolor=blue,menucolor=blue
]{hyperref}

\newcommand{\vect}[1]{\bm {#1}} 

\newcommand{\EE}{\mathcal {E}} 
\newcommand{\EllipticK}{\bm {K}} 
\newcommand{\EllipticE}{\bm {E}} 
\newcommand{\tbar}{\bar {t}} 
\newcommand{\Hopfell}{\mathfrak{L}} 

\newcommand{\re}{\tilde{r}} 
\newcommand{\se}{\tilde{s}} 

\begin{document}

\title{Finite-size effects in the short-time height distribution of the Kardar-Parisi-Zhang equation}
\author{Naftali R. Smith}
\email{naftali.smith@mail.huji.ac.il}
\author{Baruch Meerson}
\email{meerson@mail.huji.ac.il}
\affiliation{Racah Institute of Physics, Hebrew University of
Jerusalem, Jerusalem 91904, Israel}
\author{Pavel Sasorov}
\email{sasorov@gmail.com}
\affiliation{Keldysh Institute of Applied Mathematics, Moscow, 125047, Russia}

\begin{abstract}
\noindent
We use the optimal fluctuation method to evaluate the short-time probability distribution $\mathcal{P}\left(H,L,t\right)$ of height at a single point, $H=h\left(x=0,t\right)$, of the evolving Kardar-Parisi-Zhang (KPZ) interface $h\left(x,t\right)$ on a ring of length $2L$. The process starts from a flat interface. At short times typical (small) height fluctuations are unaffected by the KPZ nonlinearity and belong to the Edwards-Wilkinson universality class.
The nonlinearity, however, strongly affects the (asymmetric) tails of $\mathcal{P}(H)$. At large $L/\sqrt{t}$ the faster-decaying tail has a double structure: it is $L$-independent, $-\ln\mathcal{P}\sim\left|H\right|^{5/2}/t^{1/2}$, at intermediately large $|H|$, and $L$-dependent, $-\ln\mathcal{P}\sim \left|H\right|^{2}L/t$, at very large $|H|$.
The transition between these two regimes is sharp and, in the large $L/\sqrt{t}$ limit, behaves as a fractional-order phase transition. The transition point $H=H_{c}^{+}$ depends on $L/\sqrt{t}$. At small $L/\sqrt{t}$, the double structure of the faster tail disappears, and only the very large-$H$ tail, $-\ln\mathcal{P}\sim \left|H\right|^{2}L/t$, is observed.
The slower-decaying tail does not show any $L$-dependence at large $L/\sqrt{t}$, where it coincides with the slower tail of the GOE Tracy-Widom distribution. At small $L/\sqrt{t}$ this tail also has a double structure. The transition between the two regimes occurs at a value of height $H=H_{c}^{-}$ which depends on $L/\sqrt{t}$. At $L/\sqrt{t} \to 0$ the transition behaves as a mean-field-like second-order phase transition.  At $|H|<|H_c^{-}|$ the slower tail behaves as $-\ln\mathcal{P}\sim \left|H\right|^{2}L/t$, whereas at $|H|>|H_c^{-}|$ it coincides with the slower tail of the GOE Tracy-Widom distribution.

\end{abstract}

\maketitle

\tableofcontents
\nopagebreak

\section{Introduction}
\label{intro}

The celebrated Kardar-Parisi-Zhang (KPZ) equation \cite{KPZ} is a standard model
of non-equilibrium stochastic growth  \cite{HHZ,Barabasi,Krug,QS,S2016,Takeuchi2017}.
In $1+1$ dimension, the KPZ equation reads
\begin{equation}
\label{eq:KPZ_dimensional}
\partial_{t}h=\nu\partial_{x}^{2}h+\frac{\lambda}{2}\left(\partial_{x}h\right)^{2}+\sqrt{D}\,\xi(x,t),
\end{equation}
where $h(x,t)$ is the interface height, and $\xi(x,t)$ is a Gaussian noise with zero average
and the correlator
\begin{equation}\label{correlator}
\langle\xi(x_{1},t_{1})\xi(x_{2},t_{2})\rangle = \delta(x_{1}-x_{2})\delta(t_{1}-t_{2}).
\end{equation}

At long times the evolving KPZ interface exhibits universal scaling behavior and defines an important universality class of growth. In a sufficiently long system in  $1+1$ dimension,
the interface width grows as $t^{1/3}$, whereas the lateral correlation length grows
as $t^{2/3}$, as confirmed in
experiments with several systems belonging to the KPZ universality class \cite{Tong1994, Miettinen2005, Degawa2006, Takeuchi2010, Takeuchi2011,Takeuchi2017}.
The exponents $1/3$ and $2/3$ distinguish the KPZ universality class from the
Edwards-Wilkinson (EW) universality class which corresponds to the absence of
the nonlinear term in Eq.~(\ref{eq:KPZ_dimensional}) \citep{EW1982}.

In the recent years the focus of studies of the KPZ equation and related systems shifted towards more detailed quantities. One of them is the full distribution of the interface height $h\left(0,t\right)-h\left(0,0\right) = H$
(in a proper moving frame\footnote{\label{footnote:displacement}The evolving KPZ interface includes a systematic displacement $h_s(t)$, resulting from rectification of the noise by the nonlinearity. If the noise is delta-correlated in space, $d h_s(t)/dt$ diverges. This divergence can be regularized by introducing a small spatial correlation length of the noise. In this case $d h_s(t)/dt$ approaches, at long times, a constant which depends on the spatial correlation length of the noise \cite{Hairer,Gueudre,S2016}. We define $H(t)$ as $H(t)=h(0,t)-h_s(t)$.})
at specified point and time \citep{Corwin2012,QS,S2016,Takeuchi2017}.
Exact representations for a generating function of $\mathcal{P}(H,t)$ at arbitrary time $t > 0$ were derived, for infinite systems, for several particular cases of initial conditions and their combinations, including the flat interface $h\left(x,t=0\right)=0$ \cite{CLD}.
From the exact representations, the height distributions for typical fluctuations, in the long-time limit, were extracted. For the flat interface, $\mathcal{P}(H,t)$ converges to the Gaussian orthogonal ensemble (GOE) Tracy-Widom (TW) distribution \cite{TracyWidom1996}.
These results were confirmed in experiments with liquid-crystal turbulent fronts \cite{Takeuchi2010,Takeuchi2011,Takeuchi2017}.
Extracting the long-time and short-time \emph{tails} of  $\mathcal{P}(H,t)$ from the exact representations is a difficult task. It has been achieved for several settings \cite{DMS, SMP, DMRS, Krajenbrink2017} which do not include the flat initial condition.
This situation calls for more direct and general approaches of evaluating the tails $\mathcal{P}(H,t)$.
The need for such approaches becomes crucial for other non-equilibrium growth models where exact solutions are unavailable.

One such group of approaches is known under the names of the optimal fluctuation method (OFM) or the weak-noise theory \citep{KK2007,KK2008,KK2009,MKV,KMSparabola,Janas2016,Fogedby}. The OFM originated in condensed matter physics \cite{Halperin,Langer,Lifshitz}, see also Ref. \cite{Lifshitz1988}. The crux of the OFM is a saddle-point evaluation
of a proper path integral for the stochastic process conditioned on the large fluctuation in question and constrained by the initial and boundary conditions. This procedure brings a minimization problem which
defines an effective classical field theory. The latter can be cast into a Hamiltonian form.
Solving the Hamilton equations, one determines the optimal fluctuation, or the optimal path: the most likely history of the interface,
and the most likely realization of the noise, that give a dominant contribution to the probability of the specified large fluctuation. Once the optimal fluctuation is determined, the probability of the large fluctuation is given, up to a pre-exponential factor, by the action functional of the classical field theory. Similar methods have appeared in the studies of turbulence and turbulent transport \citep{turb1,turb2,turb3}, diffusive lattice gases (see Ref. \citep{bertini2015} for a recent review) and
stochastic reactions on lattices \citep{EK, MS2011}.

The OFM has been recently used to evaluate the single-point interface-height statistics for the KPZ equation in an infinite one-dimensional system for the flat initial condition \cite{KK2007,KK2008,KK2009,MKV},
and for other initial conditions \cite{KMSparabola, Janas2016}. Two striking features of the $\lambda H \to \infty$ tail are that (i) it holds, in the leading order,  for any time $t>0$, and (ii) it is universal for a whole class of \emph{deterministic} initial conditions, which includes all deterministic conditions studied so far \cite{KMSparabola}. In contrast to this behavior, the $\lambda H \to -\infty$ tail, as predicted by the OFM, depends on the initial conditions \cite{KMSparabola}. Analysis of exact representations \cite{Corwin2012,Sasamoto2010,Calabrese2010,Dotsenko2010,Amir2011}  for $\mathcal{P}(H,t)$ (for the sharp-wedge initial condition) shows that this tail exists for all $t>0$. At long times it is ``pushed" to progressively larger heights, so the $\lambda H \to -\infty$ tail acquires a double structure \cite{SMP}.

\begin{figure}[ht]
\includegraphics[width=0.45\textwidth,clip=]{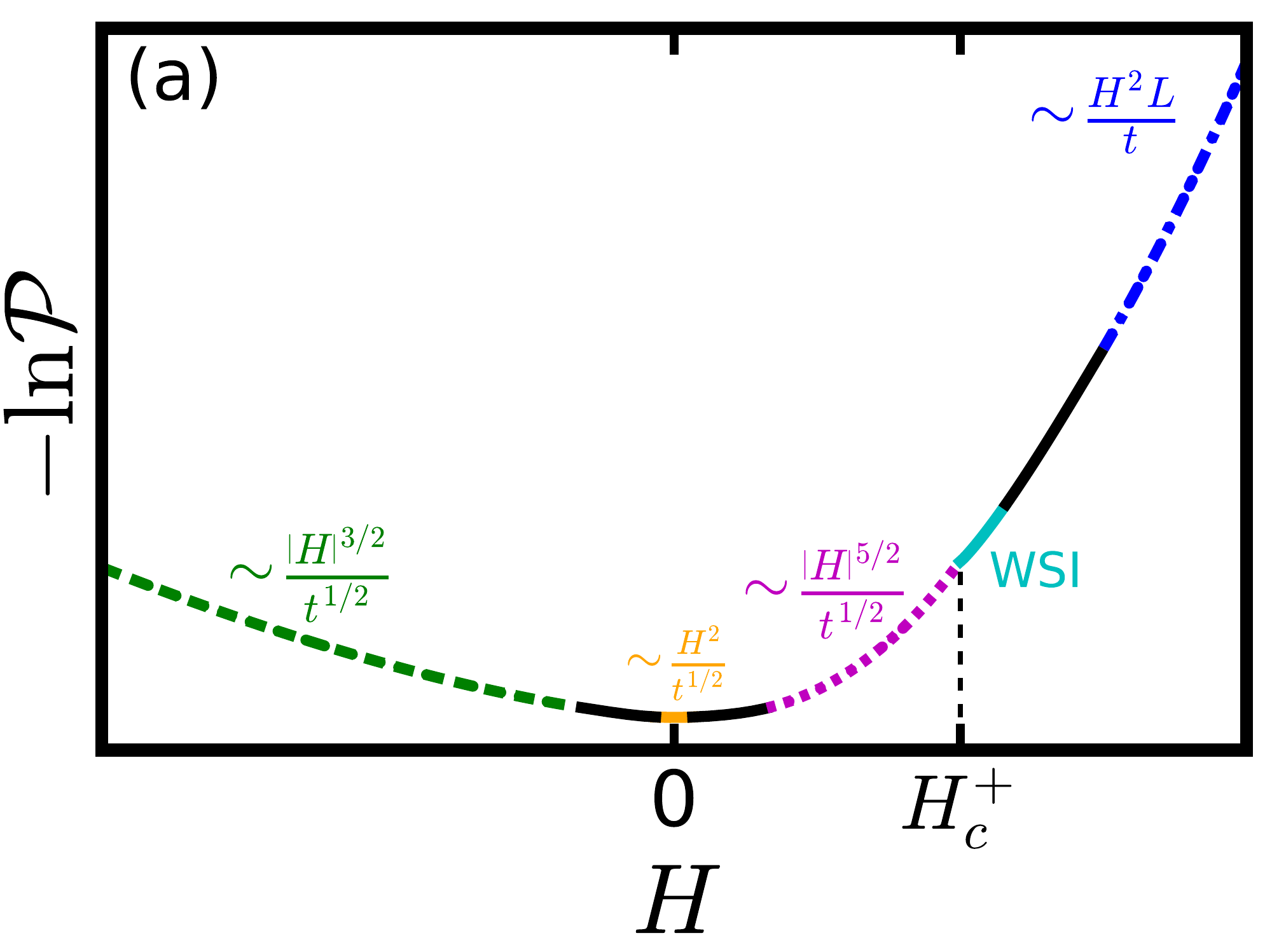}
\includegraphics[width=0.45\textwidth,clip=]{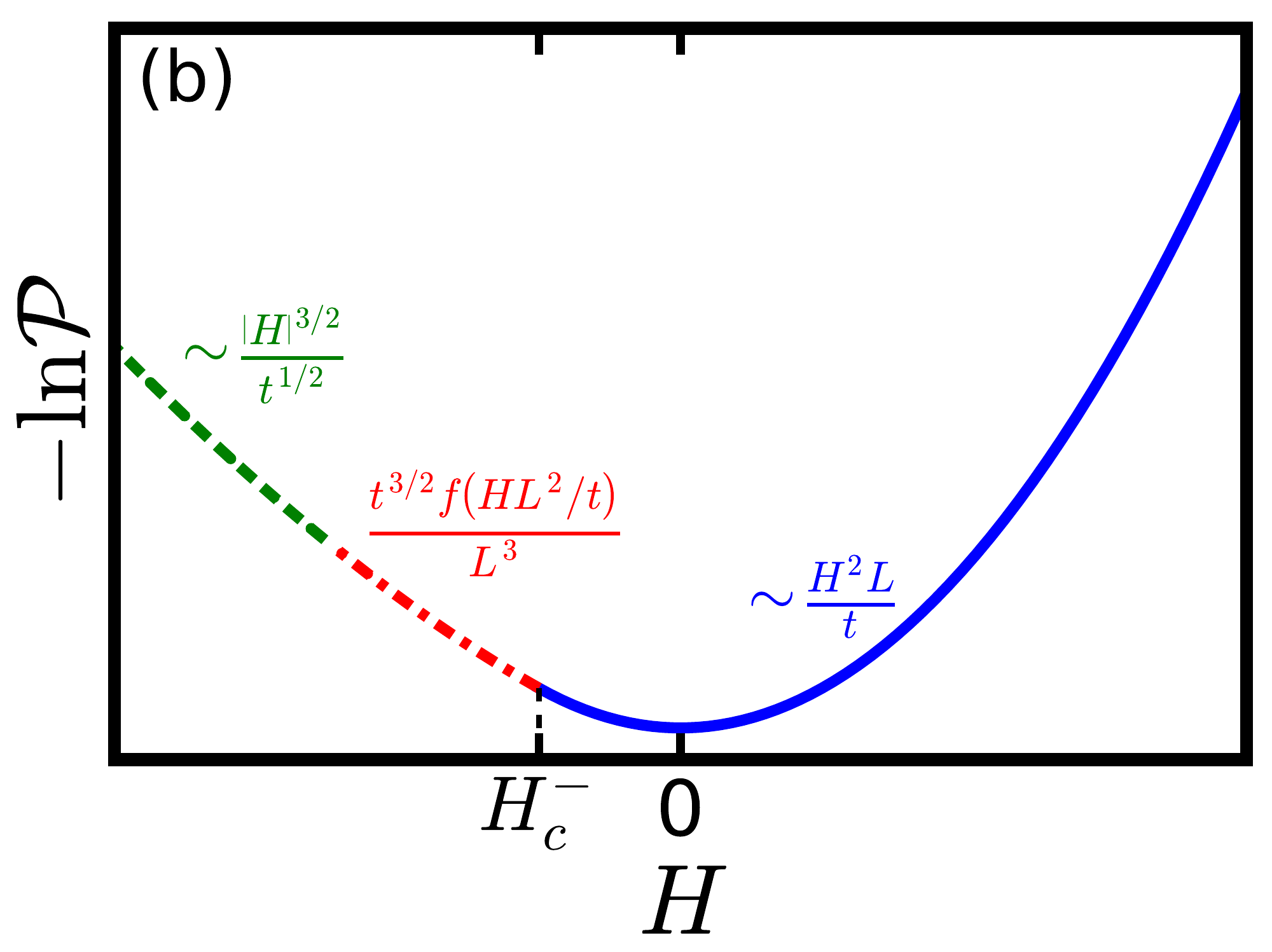}
\caption{A schematic plot of the logarithm of the short-time ($t\ll\nu^{5} \! /D^{2}\lambda^{4}$) height distribution at fixed $L$ and $t$, for $L \gg \sqrt{\nu t}$ (a) and $L \ll \sqrt{\nu t}$ (b). See main text for details. WSI denotes the weakly-supercritical inviscid region,  discussed in Sec. \ref{section:small_hole}.}
\label{fig:S_schematic}
\end{figure}

What happens if the one-dimensional system has a finite size? Here we address this question in the weak-noise (or short-time) regime, when the typical fluctuations of height are described by the EW equation. Our focus will be on the \emph{tails} of $\mathcal{P}(H,L,t)$ which, for finite systems, are presently unknown. We employ the OFM to determine these tails for the KPZ equation on a ring of length $2L$. As we will show,  the finite-$L$ effects are quite interesting, and we identify  the ``phase diagram" of the system on the plane $(L/\sqrt{\nu t}, \;\lambda H/\nu)$ which describes different regimes of the optimal fluctuations and of the ensuing height statistics.
Figure~\ref{fig:S_schematic} presents, in a schematic form, the scaling behavior of the height distribution $\mathcal{P}\left(H,L,t\right)$ up to numerical factors.
In a proper moving frame\footnote{See footnote \ref{footnote:displacement}.} and at short times $t\ll\nu^{5}/D^{2}\lambda^{4}$, the body of $\mathcal{P}\left(H,L,t\right)$ is Gaussian.  The distribution tails, however,  are strongly asymmetric.
At $L/\sqrt{\nu t} \gg 1$ the $\lambda H \to -\infty$ tail has a double structure.
At intermediately large $H$ its behavior $-\ln\mathcal{P}\sim\left|H\right|^{5/2}/t^{1/2}$ is independent of $L$, whereas at very large $H$ it is $L$-dependent, $-\ln\mathcal{P}\sim \left|H\right|^{2}L/t$. The transition between the two regimes is sharp: in the large $L/\sqrt{\nu t}$ limit it has a character of a fractional-order phase transition, which occurs at an $L/\sqrt{\nu t}$-dependent critical value $H=H_c^+$. Interestingly, for very large $\lambda H>0$ there are no finite-$L$ effects at all, and this tail coincides with the slower-decaying tail of
the GOE Tracy-Widom distribution, $-\ln\mathcal{P}\sim \left|H\right|^{3/2} \!/\, t^{1/2}$.

At $L/\sqrt{\nu t} \ll 1$, the double structure of the $\lambda H \to -\infty$ tail disappears. Only the very large-$H$ tail is observed. In contrast, the $\lambda H \to +\infty$ tail has a double structure: we find a sharp transition of the large deviation function around
$H=H_{c}^{-}$ which depends on $L/\sqrt{\nu t}$.  Here
the transition becomes a singularity of the large deviation function which has the character of a (mean-field-like) second-order phase transition. For $|H|<|H_c^{-}|$, the optimal interface is spatially uniform leading to $-\ln\mathcal{P}\sim \left|H\right|^{2}L/t$. For $|H|>|H_c^{-}|$, the optimal interface is a traveling front and, at very large $|H|$, the tail approaches the slower-decaying
tail of the GOE Tracy-Widom distribution,
$-\ln\mathcal{P}\sim \left|H\right|^{3/2} \!/\, t^{1/2}$. In the limit of $L/\sqrt{t}\to0$,
the height distribution scales as $-\ln\mathcal{P} \! \sim \! (t^{3/2}/L^{3}) f\left(HL^{2}/t\right)$, and we calculate the scaling function $f$.

The derivations, the numerical factors, the optimal paths and additional details are presented in the relevant sections of the paper and in the Appendices. In Sec. \ref{Scaling} we present the OFM formulation of the problem, discuss the spatially uniform solution to the OFM equations and, through simple arguments, predict many of the results of this paper up to numerical coefficients. In Sec. \ref{sec:typical} we derive,  for completeness, the Gaussian distribution of typical fluctuations with account of the finite-$L$ effects. In Sec. \ref{sec:negative_tail} we determine the optimal path and the height distribution for large  $\lambda H > 0$, including the $\lambda H\to +\infty$ tail.
Section \ref{sec:positive_tail} deals with the $\lambda H \to -\infty$ tail.
We briefly summarize and discuss our results in Sec. \ref{disc}. Some of the technical details are relegated to Appendices.

\section{OFM on a ring}
\label{Scaling}

\subsection{Governing equations and constraints}
Let us introduce the observation time $T$ at which the interface height is measured, $H=h\left(0,T\right)-h\left(0,0\right)$.
Rescaling the variables,  $x/\sqrt{\nu T}\to x$, $t/T \to t$, $\left|\lambda\right|h/\nu\to h$, we  bring Eq.~(\ref{eq:KPZ_dimensional}) to the dimensionless form \citep{MKV}
\begin{equation}
\label{eq:KPZ_dimensionless}
\partial_{t}h=\partial_{x}^{2}h-\frac{1}{2}\left(\partial_{x}h\right)^{2}+\sqrt{\epsilon} \, \xi\left(x,t\right),
\end{equation}
where $\epsilon=D\lambda^{2}\sqrt{T}/\nu^{5/2}$ is the rescaled strength of the noise, and we assume, without loss of generality, that $\lambda<0$
\footnote{Changing $\lambda$ to $-\lambda$ is  equivalent to changing $h$ to $-h$.}.
The rescaled system length is $2\ell = 2L/\sqrt{\nu T}$.  We assume a flat initial condition,
\begin{equation}
\label{eq:flat_initial_cond}
h\left(x,t=0\right)=0.
\end{equation}
Let the interface height $H$ (rescaled by $\nu / |\lambda|$) be reached at $x = 0$. In the weak-noise (or short-time) limit, $\epsilon \ll 1$,  one can use a saddle-point evaluation of the proper path integral of Eq.~(\ref{eq:KPZ_dimensionless}). This procedure leads to a minimization problem for the effective action \citep{KK2007, Fogedby, MKV}.
The minimal action is found from the optimal history $h\left(x,t\right)$ of the height profile and its canonically conjugate ``momentum'' field $\rho\left(x,t\right)$. The fields $h$ and $\rho$ satisfy the equations \citep{Fogedby, KK2007, MKV}
\begin{eqnarray}
\label{eq:OFM_h}
\partial_{t}h	&=&	\frac{\delta\mathcal{H}}{\delta\rho}=\partial_{x}^{2}h-\frac{1}{2}\left(\partial_{x}h\right)^{2}+\rho,\\
\label{eq:OFM_rho}
\partial_{t}\rho	&=&	-\frac{\delta\mathcal{H}}{\delta h}=-\partial_{x}^{2}\rho-\partial_{x}\left(\rho\partial_{x}h\right),
\end{eqnarray}
where $\mathcal{H}=\int_{-\ell}^{\ell}w\,dx$ is the Hamiltonian, and $w\left(x,t\right)=\rho\left[\partial_{x}^{2}h-\left(1/2\right)\left(\partial_{x}h\right)^{2}+\rho/2\right]$.
The initial condition is Eq.~(\ref{eq:flat_initial_cond}). A ring of length $2\ell$ imposes the periodic boundary conditions
\begin{equation}
\label{boundary_cond}
h\left(x+2\ell,t\right)=h\left(x,t\right),\quad\rho\left(x+2\ell,t\right)=\rho\left(x,t\right).
\end{equation}
The constraint $h\left(0,1\right)=H$ can be taken into account by introducing a Lagrange multiplier into the action functional, giving rise to the boundary condition at $t=1$:
\begin{equation}
\label{eq:delta_initial_cond_rho}
\rho\left(x,1\right)=\Lambda\delta(x),
 \end{equation}
where $\Lambda$ is ultimately determined by $H$ and $\ell$.
For completeness, we present a derivation of Eqs.~(\ref{eq:OFM_h}) and (\ref{eq:OFM_rho}), and of the boundary condition (\ref{eq:delta_initial_cond_rho}), in
Appendix \ref{Appendix:OFMderivation}. After solving the OFM problem, the height distribution is found, up to a pre-exponential factor, by evaluating
\begin{equation}
\label{eq:action_scaling}
-\ln\mathcal{P}\left(H,L,T\right)\simeq\frac{S\left(\frac{\left|\lambda\right|H}{\nu},\frac{L}{\sqrt{\nu T}}\right)}{\epsilon}=\frac{\nu^{5/2}}{D\lambda^{2}\sqrt{T}}S\left(\frac{\left|\lambda\right|H}{\nu},\frac{L}{\sqrt{\nu T}}\right),
\end{equation}
where the rescaled action $S$ is given by
\begin{equation}
\label{eq:action_definition}
S=\int_{0}^{1}dt\int_{-\ell}^{\ell}dx
\left(\rho\partial_{t}h-w\right)=\frac{1}{2}\int_{0}^{1}dt\int_{-\ell}^{\ell}dx\rho^{2}\left(x,t\right).
\end{equation}
The rescaled action $S$ can be interpreted as the large deviation function of height at short times. In the absence of exact solution for arbitrary $H$ and $L$, we will obtain approximate solutions in different limits.

Motivated by previous work on infinite systems \citep{KK2009, MKV, KMSparabola} and by our numerical calculations (see below), we assume throughout this paper that the optimal paths of the system
preserve the mirror symmetry $x \leftrightarrow -x$:
\begin{equation}
\label{eq:h_symmetric}
h\left(x,t\right)=h\left(-x,t\right), \quad \rho\left(x,t\right)=\rho\left(-x,t\right).
\end{equation}
Previously this symmetry was only seen to be broken for random initial conditions \cite{Janas2016}, but never for any (mirror-symmetric) deterministic  initial condition.

\subsection{Spatially uniform solution}
\label{CDSL}
Equations~(\ref{eq:OFM_h}) and~(\ref{eq:OFM_rho}) have the obvious spatially uniform solution
\begin{equation}
\label{eq:uniform_solution}
h\left(x,t\right)=Ht, \quad \rho(x,t)=H.
\end{equation}
This solution would not be allowed in an infinite system, as it would give an infinite action and a zero probability. Plugging it into Eq.~(\ref{eq:action_definition}) gives the action
\begin{equation}
\label{eq:action_uniform_solution}
S=H^{2} \ell,
\end{equation}
which corresponds to a probability distribution
\begin{equation}
\label{eq:height_distribution_uniform}
-\ln\mathcal{P}\left(H,L,T\right) \simeq \frac{H^{2}L}{DT},
\end{equation}
which is Gaussian in $H$ and independent of $\nu$ and $\lambda$. The uniform solution satisfies the initial condition~(\ref{eq:flat_initial_cond}) and the constraint $h(0,1)=H$, but it does not obey the boundary condition~(\ref{eq:delta_initial_cond_rho}), and so it is not optimal.  Nevertheless, as we will see shortly, for typical fluctuations in short systems $\ell \ll 1$, and for the $H\to+\infty$ tail at any $\ell$, the action is indeed given, in the leading order, by Eq.~(\ref{eq:action_uniform_solution}). In these regimes, the correct solution to the OFM problem is nearly uniform for most of the dynamics, except for a short temporal boundary layer near the end of the time interval $0 < t < T$, where $\rho$ rapidly adapts to the boundary condition~(\ref{eq:delta_initial_cond_rho}).

As to be expected, the action (\ref{eq:action_uniform_solution}) tends to zero, and the variance of the height distribution diverges,
in the limit of $\ell\to 0$. Indeed, at very long times (or for very short systems) the height distribution is described by the stationary measure of the KPZ equation \cite{HHZ, Barabasi}, so that
\begin{equation}
\label{eq:free_energy}
-\ln\mathcal{P}\left(H,L\right) \simeq \frac{1}{2 \epsilon}\int_{-\ell}^{\ell}\left(\partial_{x}h\right)^{2}dx.
\end{equation}
The right hand side is the equilibrium free energy of the KPZ interface in one dimension \cite{HHZ,Barabasi}. Minimizing it under the constraint $h\left(x=0\right)=H$, one arrives at the uniform profile $h\left(x\right)=H$,
a zero action and an infinite variance of the height distribution.

As mentioned above, Eq.~(\ref{eq:height_distribution_uniform}) presents a correct leading-order asymptotic for typical fluctuations in short systems $\ell \ll 1$, and for the $H\to+\infty$ tail.  In other regimes of parameters (for example, for typical fluctuations in large systems, $\ell \gg 1$), the uniform solution cannot adapt to the boundary condition~(\ref{eq:delta_initial_cond_rho}), so it is not a valid approximate solution.

\subsection{Simple estimates}

Many of the results of this work can be foreseen, up to numerical coefficients, from simple scaling arguments and
order-of-magnitude estimates. Let us start with an infinite system. One of the striking results there is the strong asymmetry between the positive and negative short-time tails of the height distribution $\mathcal{P}\left(H,t\right)$ \citep{KK2007,MKV}. This asymmetry comes from the very different character of the optimal path which, in its turn, is explained by the very different role of diffusion. For the positive tail,
$H\gg 1$, the KPZ nonlinearity (with the diffusion neglected) leads to a cusp formation. The optimal realization of noise $\rho(x,t)$ must, on the one hand, cause the interface to reach a large height at $x=0$ and, on the other hand, delay the cusp formation until $t=1$. As a result, the characteristic spatial scale of $\rho(x,t)$, which we will call $\delta$, during most of the dynamics must be much larger than the diffusion length scale, which is of order 1. In this regime the diffusion can be neglected altogether. Balancing the remaining terms in Eq.~(\ref{eq:OFM_h}), $H\sim H^2/\delta^2\sim \rho$, we obtain the scalings $\rho \sim H$ and $\delta \sim \sqrt{H}$. The resulting action~(\ref{eq:action_definition}) (with $\ell =\infty$) scales as $S\sim \rho^2 \delta \sim H^{5/2}$.

In the negative tail, $-H\gg 1$, the KPZ nonlinearity does not lead to a cusp even in the absence of diffusion. Therefore, a minimum action
is achieved when the optimal realization of noise $\rho$ is strongly localized. Here $\delta$,  the characteristic spatial scale of $\rho$,  is determined by a competition between the nonlinearity and diffusion. In addition, $\rho$ is independent of time for most of the dynamics \citep{KK2007,MKV}.
Balancing the terms in Eqs.~(\ref{eq:OFM_h}) and~(\ref{eq:OFM_rho}), we again obtain $\rho \sim H$, but now $\delta \sim 1/\sqrt{|H|}$. As a result, the action scales as $S\sim \rho^2 \delta \sim |H|^{3/2}$.

Now let us return to the ring problem and to the fully dimensional variables.  In the limit of $L \to 0$,  for fixed $H$ and $T$, the interface is spatially uniform, leading to Eq.~(\ref{eq:height_distribution_uniform}). For small fluctuations the KPZ nonlinearity is negligible, and $\mathcal{P}(H,L,T)$ cannot depend on $\lambda$. Then it follows from Eq.~(\ref{eq:action_scaling}) that
\begin{equation}
\label{eq:handwaving_EW}
-\ln\mathcal{P}\left(H,L,T\right)\simeq\frac{H^{2}\sqrt{\nu}}{D\sqrt{T}}G\left(\frac{L}{\sqrt{\nu T}}\right)
\end{equation}
with some scaling function $G$. As to be expected, small fluctuations of $H$ are normally distributed. Finite-size effects come into play when the system size $2L$ is comparable with, or shorter than the diffusion length scale $\sqrt{\nu T}$.

For large negative $\lambda H$, the diffusion is negligible \citep{KK2007,KK2009,MKV,KMSparabola, Janas2016}, so Eq.~(\ref{eq:action_scaling}) implies
\begin{equation}
\label{eq:handwaving_inviscid}
-\ln\mathcal{P}\left(H,L,T\right)\simeq
\frac{\sqrt{\left|\lambda\right|}\,H^{5/2}}{D\sqrt{T}}s\left(\frac{L}{\sqrt{\left|\lambda\right|HT}}\right).
\end{equation}
Here finite-size effects become important when the system size $2L$ is comparable with, or shorter than the ``inviscid" length scale $\delta$ which, in the dimensional variables, is $\delta \sim \sqrt{\left|\lambda\right|HT}$.
For $L \ll \delta$ we expect to reproduce Eq.~(\ref{eq:height_distribution_uniform}).
This implies a Gaussian tail at $\lambda H \to -\infty$. For  $L \gg \delta$ $L$  must drop out of the result, yielding the infinite-system result $-\ln\mathcal{P} \sim  H^{5/2}/T^{1/2}$.

For large positive $\lambda H$, $\rho(x,t)$ and $\partial_{t}h\left(0,t\right)$ are $t$-independent of time for most of the dynamics. This important property implies a scaling relation
\begin{equation}
\label{eq:homogeneous_HT_scaling}
-\ln\mathcal{P}\left(H,L,T\right)\simeq T\tilde{f}\left(\frac{H}{T},L\right).
\end{equation}
Combining Eqs.~(\ref{eq:action_scaling}) and~(\ref{eq:homogeneous_HT_scaling}), we obtain
\begin{equation}
\label{eq:handwaving_positive_Hscaling}
-\ln\mathcal{P}\left(H,L,T\right)\simeq\frac{\nu^{4}T}{D\lambda^{2}L^{3}}f\left(\frac{\left|\lambda\right|HL^{2}}{\nu^{2}T}\right).
\end{equation}
Here finite-size effects are expected to show up when $2L$ is comparable with, or shorter than $\nu\sqrt{T/\left(\left|\lambda\right|H\right)}$ which is nothing but the corresponding $\delta$ for an infinite system, expressed in the dimensional variables. For $L\ll \delta$ Eq.~(\ref{eq:handwaving_positive_Hscaling}) should coincide with Eq.~(\ref{eq:height_distribution_uniform}). For $L\gg \delta$ $L$ drops out of the result implying $-\ln\mathcal{P} \sim \left|H\right|^{3/2}\!/T^{1/2}$.

In the following sections, we derive these scaling behaviors more rigorously, calculate the scaling functions $G$ and $f$ exactly and calculate $s$ in certain limits. No less important, we will uncover phase-transition-like behaviors which are missed by simple scaling analyses like these.

\section{Edwards-Wilkinson regime}
\label{sec:typical}
\begin{figure}[ht]
\includegraphics[width=0.5\textwidth,clip=]{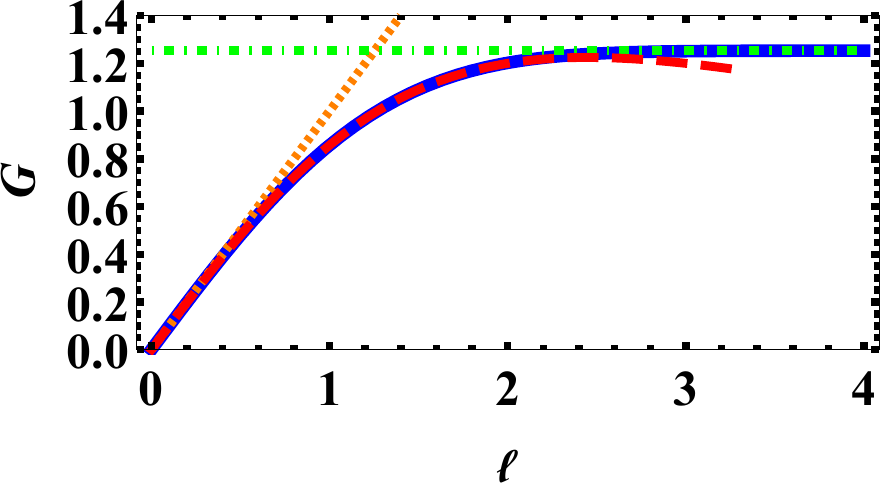}
\caption{The scaling function $G(\ell)$~(\ref{eq:Stilde_L}) (solid line) which describes the Gaussian distribution of typical fluctuations of the interface height. The small $\ell$ (dashed) and large $\ell$ (dot-dashed) asymptotics are shown, see Eqs.~(\ref{eq:S_small_L_linear_theory}) and~(\ref{eq:S_large_L_linear_theory}), as well as the action evaluated on the uniform solution~(\ref{eq:action_uniform_solution}) (dotted).}
\label{fig:Stilde}
\end{figure}

For sufficiently small heights $H$ we can solve, at arbitrary $L/\sqrt{T}$, the OFM equations perturbatively with respect to $H$ or $\Lambda$ \citep{MKV}.
We expand
\begin{eqnarray}
\label{eq:lineartization_h}
h\left(x,t\right)	        &=&	\Lambda h_{1}\left(x,t\right)+\Lambda^{2}h_{2}\left(x,t\right)+\dots,\\
\label{eq:lineartization_rho}
\rho\left(x,t\right)	&=&	\Lambda\rho_{1}\left(x,t\right)+\Lambda^{2}\rho_{2}\left(x,t\right)+\dots.
\end{eqnarray}
Correspondingly, $S\left(\Lambda\right)=\Lambda^{2}S_{1}+\Lambda^{3}S_{2}+\dots$.
In the leading order Eqs.~(\ref{eq:OFM_h}) and~(\ref{eq:OFM_rho}) become
\begin{eqnarray}
\label{eq:OFM_EW_h}
\partial_{t}h_{1}     &=& \; \partial_{x}^{2}h_{1}+\rho_{1}, \\
\label{eq:OFM_EW_rho}
\partial_{t}\rho_{1} &=& -\partial_{x}^{2}\rho_{1}.
 \end{eqnarray}
These linear equations correspond to the OFM theory for the Edwards-Wilkinson (EW) equation \cite{EW1982}
\begin{equation}
\label{eq:EW}
\partial_{t}h=\nu\partial_{x}^{2}h+\sqrt{D}\,\xi(x,t).
\end{equation}
Solving Eq.~(\ref{eq:OFM_EW_rho}) backward in time with initial condition $\rho_{1}\left(x,t=1\right)=\delta\left(x\right)$, we obtain:
 \begin{equation}
 \label{eq:EW_rho1sol}
\rho_{1}\left(x,t\right)=\frac{1}{2\ell}+\frac{1}{\ell}\sum_{n=1}^{\infty}\cos\left(\frac{\pi nx}{\ell}\right)\exp\left[-\frac{\pi^{2}n^{2}\left(1-t\right)}{\ell^{2}}\right].
  \end{equation}
As a result,
\begin{equation}
\label{eq:EW_S1sol}
S_{1}=\frac{1}{2}\int_{0}^{1}dt\int_{- \ell }^{ \ell }dx \, \rho_{1}^{2}\left(x,t\right)=\frac{1}{4 \ell }+\frac{ \ell }{4\pi^{2}}\sum_{n=1}^{\infty}\frac{1}{n^{2}}\left[1-\exp\left(-\frac{2\pi^{2}n^{2}}{ \ell ^{2}}\right)\right].
\end{equation}
To express $\Lambda$ in terms of $H$ and $\ell$, we solve
Eq.~(\ref{eq:OFM_EW_h}) with the forcing term $\rho_{1}$ from Eq.~(\ref{eq:EW_rho1sol}) and the initial condition~(\ref{eq:flat_initial_cond}). After standard algebra, the solution is
 \begin{equation}
 \label{eq:EW_h1sol}
h_{1}\left(x,t\right)=\frac{t}{2\ell}+\frac{\ell}{\pi^{2}}\sum_{n=1}^{\infty}\frac{1}{n^{2}}\exp\left(-\frac{\pi^{2}n^{2}}{\ell^{2}}\right)\sinh\left(\frac{\pi^{2}n^{2}t}{\ell^{2}}\right)\cos\left(\frac{\pi nx}{\ell}\right),
\end{equation}
from which we obtain
 \begin{equation}
  \label{eq:EW_H_Lambda}
 H=\Lambda h_{1}\left(x=0,t=1\right)=\Lambda\left\{ \frac{1}{2 \ell }+\frac{ \ell }{2\pi^{2}}\sum_{n=1}^{\infty}\frac{1}{n^{2}}\left[1-\exp\left(-\frac{2\pi^{2}n^{2}}{ \ell ^{2}}\right)\right]\right\}.
\end{equation}
From Eqs.~(\ref{eq:EW_S1sol}) and~(\ref{eq:EW_H_Lambda}) we obtain the Gaussian action
 \begin{equation}
 \label{eq:Stilde_L}
S=H^{2}G\left( \ell \right), \qquad G\left( \ell \right)=\frac{1}{2}\left\{ \frac{1}{2 \ell }+\frac{ \ell }{2\pi^{2}}\sum_{n=1}^{\infty}\frac{1}{n^{2}}\left[1-\exp\left(-\frac{2\pi^{2}n^{2}}{ \ell ^{2}}\right)\right]\right\} ^{-1}
 \end{equation}
which is consistent with the simple estimate~(\ref{eq:handwaving_EW}).
 For $\ell\ll1$, we neglect the exponential terms in Eq.~(\ref{eq:Stilde_L}) and obtain
\begin{equation}
 \label{eq:S_small_L_linear_theory}
S\simeq\frac{H^{2}\ell}{1+\frac{\ell^{2}}{6}}.
\end{equation}
This $S$ is smaller than the action along the uniform path~(\ref{eq:action_uniform_solution}) and approaches the latter
as $\ell\to 0$.
In the opposite limit $\ell\gg1$,
the sum in Eq.~(\ref{eq:Stilde_L}) can be approximated by an integral, leading to the well known result \citep{Krug2,MKV}
\begin{equation}
 \label{eq:S_large_L_linear_theory}
S \simeq \sqrt{\frac{\pi}{2}}\,H^2 .
\end{equation}
Equation~(\ref{eq:S_large_L_linear_theory}) is valid up to corrections which are exponentially small in $\ell$. This is because algebraic corrections all vanish by virtue of the Euler-Maclaurin formula \citep{EulerMaclaurin}.  The graph of $G\left( \ell \right)$ from Eq.~(\ref{eq:Stilde_L}), alongside with its small- and large-$\ell$ asymptotes, are shown in Fig. \ref{fig:Stilde}.

The optimal interface is
\begin{equation}
 \label{eq:EW_hsol}
h\left(x,t\right)\simeq\Lambda h_{1}\left(x,t\right)=2HG\left(\ell\right)\left[\frac{t}{2\ell}+\frac{\ell}{\pi^{2}}\sum_{n=1}^{\infty}\frac{1}{n^{2}}\exp\left(-\frac{\pi^{2}n^{2}}{\ell^{2}}\right)\sinh\left(\frac{\pi^{2}n^{2}t}{\ell^{2}}\right)\cos\left(\frac{\pi nx}{\ell}\right)\right].
\end{equation}
One can show that, at $\ell \gg 1$, Eq.~(\ref{eq:EW_hsol}) coincides with the infinite-system result \cite{MKV}
\footnote{A simple way to see this is by differentiating Eq.~(\ref{eq:EW_hsol}) twice. The infinite sum can then be replaced by an integral, and the resulting $\partial_{x}^{2}h$ takes the form of a difference of two Gaussians in $x$. It is then straightforward to obtain $h(x,t)$ by twice integrating in $x$.}.
At $\ell \ll 1$ the first term in Eq.~(\ref{eq:EW_hsol}) describes the uniform solution (\ref{eq:uniform_solution}). The remaining series gives the leading-order correction. At $t=1$, this series can be evaluated, and one obtains
\begin{equation}
 \label{eq:EW_hsol_t_1}
h\left(x,t=1\right) \simeq H\left(1-\frac{\left|x\right|\ell}{2}+\frac{x^{2}}{4}\right).
\end{equation}
There is a corner singularity at $x=0$ in Eq.~(\ref{eq:EW_hsol_t_1}), as for infinite systems \cite{MKV}.

The linear (EW) approximation to the KPZ equation is valid when the diffusion terms in the OFM equations~(\ref{eq:OFM_h}) and~(\ref{eq:OFM_rho}) are much larger than the nonlinear terms. Evaluating the diffusion and nonlinear terms on the leading order term, $\Lambda h_{1}$, of the optimal height profile at $t=1$, one can show from Eq.~(\ref{eq:EW_h1sol}) that the condition is
$\left|H\right|\ll\max\left\{ 1,\ell^{-2}\right\}$, or
 \begin{equation}
\frac{\left|\lambda H\right|}{\nu}\ll\max\left\{ 1,\,\frac{\nu T}{L^{2}}\right\}
\end{equation}
in the dimensional variables.

A linear problem similar to that considered in this section has been recently studied in Ref.  \cite{Gross}. The author used the OFM to study the EW equation in finite systems and calculated the one-point statistics of the \emph{relative} height, defined (in our notation) as
\begin{equation*}
h\left(x=0,t=T\right)-\frac{1}{2L}\int_{-L}^{L}h\left(x,t=T\right)dx.
\end{equation*}

\section{Large positive $\lambda H$}
\label{sec:negative_tail}

In this section we determine the height distribution for large positive $\lambda H$. In the small $\ell$ limit, we uncover an abrupt change of the large deviation function $S$ around a critical height $H=H_{c}^{-} (\ell)$. At $\ell \to 0$ this abrupt change has the character of a second-order dynamical phase transition.

\subsection{Traveling $h$-front solution and a second-order phase transition at $L/\sqrt{T} \to 0$}

\begin{figure}[ht]
\includegraphics[width=0.5\textwidth,clip=]{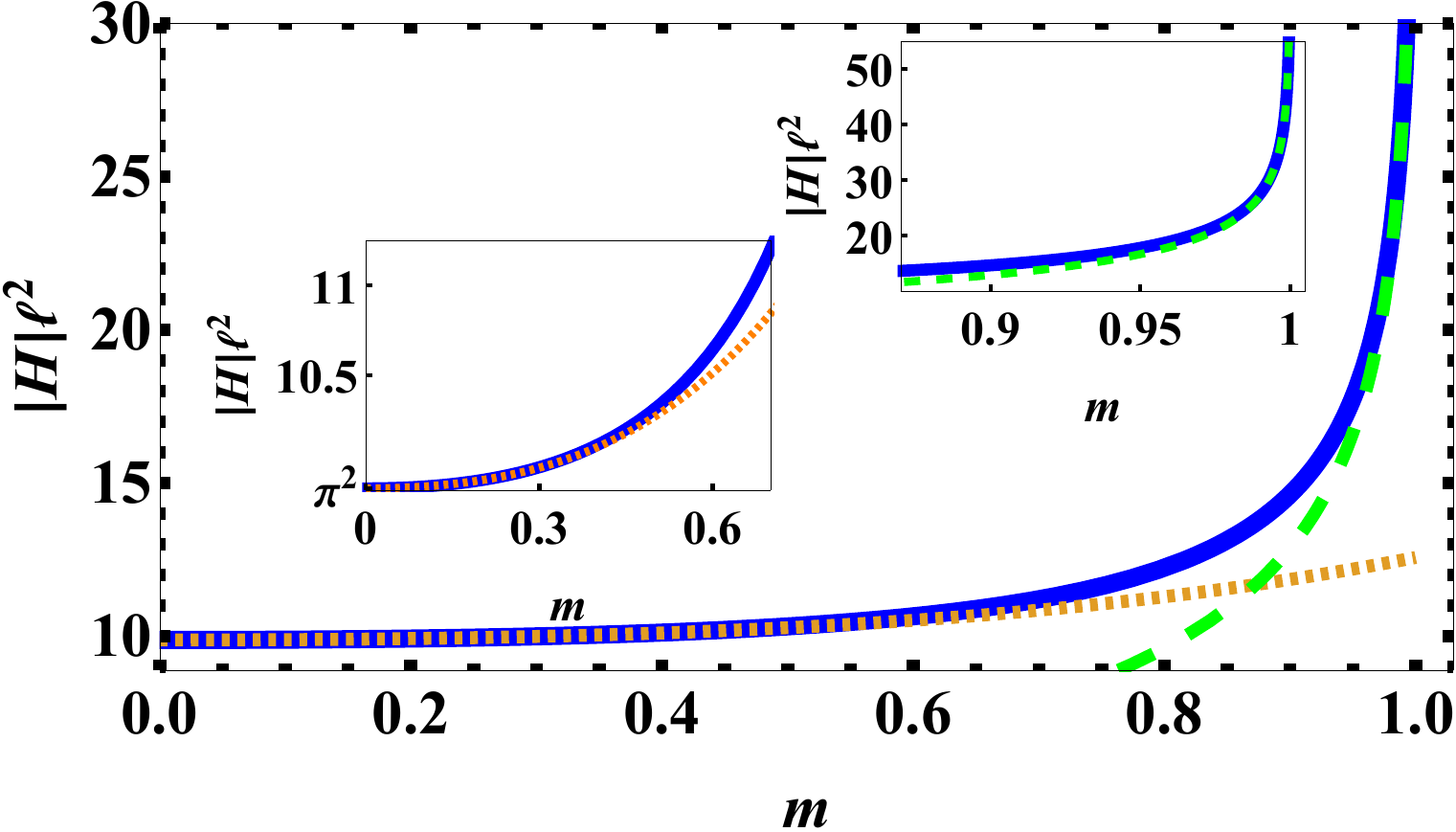}
\caption{$|H|\ell^{2}$ as a function of $ m $ for the traveling $h$-front solution, see Eq.~(\ref{eq:c_as_function_of_kappa}). Shown are the exact expression (solid) and the $ m  \to 0^+$ (dotted) and $ m  \to 1^-$ (dashed) asymptotics, which correspond to the weakly-supercritical~(\ref{eq:weakly_supercritical_Hl_squared}) and strongly-supercritical~(\ref{eq:strongly_supercritical_Hl_squared}) limits, respectively.
Insets: the regimes $ m  \ll 1$ and $1- m  \ll 1$.}
\label{fig:HLsquared_vs_kappa}
\end{figure}

In an infinite system the optimal path of the system for large negative $H$ is given,
in the leading order, by a stationary soliton of $\rho(x)$ which drives two outgoing symmetric
$h$-ramps \cite{KK2007,MKV}.  At large distances these ramps match with the unperturbed interface $h=0$. A key element of this construction is described by the ansatz
\begin{eqnarray}
h\left(x,t\right)    &=&  h_{0}\left(x\right)-ct, \label{eq:h0def}\\
\rho\left(x,t\right) &=& \rho_{0}\left(x\right). \label{eq:rho0def}
\end{eqnarray}
It is natural to call the solution~(\ref{eq:h0def}) for $h(x,t)$ a traveling $h$-front. As we found here, the ansatz~(\ref{eq:h0def}) and (\ref{eq:rho0def}) plays a key role in the finite system as well, and
generalizes the infinite-system solution \cite{KK2007,MKV}. Plugging Eqs.~(\ref{eq:h0def}) and~(\ref{eq:rho0def}) into Eqs.~(\ref{eq:OFM_h}) and~(\ref{eq:OFM_rho}) and integrating Eq.~(\ref{eq:OFM_rho}) once gives:
\begin{eqnarray}
\label{eq:OFM_TW_V0}
V_{0}'-\frac{1}{2}V_{0}^{2}+\rho_{0}	 &=&	-c,\\
\label{eq:OFM_TW_rho0}
\rho_{0}'+\rho_{0}V_{0}	&=&	C_1,
\end{eqnarray}
where $V_{0}\left(x\right)=h_{0}'\left(x\right)$ is the slope of the traveling $h$-profile, and $C_1$ is a constant. Assuming the mirror symmetry~(\ref{eq:h_symmetric}), $V_{0}$ and $\rho_{0}'$  must both vanish at $x=0$ and at $x=\pm \ell $, hence $C_1 = 0$. The resulting equations are Hamiltonian \cite{MKV},
$V_{0}'=\partial_{\rho_{0}} \mathfrak{h}$ and $\rho_{0}'=-\partial_{V_{0}} \mathfrak{h}$, with the Hamiltonian
\begin{equation}
\label{eq:hamiltonian_hh_def}
\mathfrak{h}\left(V_{0},\rho_{0}\right)=\frac{\rho_{0}}{2}\left(V_{0}^{2}-\rho_{0}-2c\right).
\end{equation}
They can be solved exactly, see Appendix \ref{Appendix:TW}. The solution for $\rho_0(x)$ is
\begin{equation}
\label{eq:rho0_negative_tail_exact}
\rho_{0}\left(x\right) = -\left[\frac{2\EllipticK\left( m \right)}{\ell}\right]^{2}\text{dn}^{2}\left[\frac{\EllipticK\left( m \right)x}{\ell},\, m \right],
\end{equation}
where $\EllipticK( m )$  is the complete elliptic integral of the first kind \citep{elliptic_integrals} and $\text{dn}(u,  m )$ is a Jacobi elliptic function \citep{elliptic_functions}. The elliptic modulus $0< m <1$  is related to $H$ and $\ell$ via the equation
\begin{equation}
\label{eq:c_as_function_of_kappa}
c=\frac{2\left(2- m \right) \EllipticK^2\left( m \right)}{\ell^2}\simeq  \left|H\right|,
\end{equation}
where we have already imposed the constraint $h(0,1)=H$.  It follows from Eq.~(\ref{eq:c_as_function_of_kappa}) that $|H|\ell^2$ is a monotone increasing function of $ m $, see Fig. \ref{fig:HLsquared_vs_kappa}.  $ m =0$ yields a critical value of $H$:
\begin{equation}
\label{eq:Hcminus_def}
\left|H_{c}^{-}\right|=\frac{\pi^{2}}{\ell^{2}}.
\end{equation}
The nontrivial traveling $h$-front solution exists only for supercritical heights, $\left|H\right|>\left|H_{c}^{-}\right|$. For subcritical heights, $\left|H\right|<\left|H_{c}^{-}\right|$, the traveling $h$-profile becomes flat. To obey the conditions  $h(0,x)=0$ and $h(0,1)=H$, one chooses $h_0(x)=0$ and $c=-H$. As a result,  the subcritical solution coincides with the spatially uniform solution~(\ref{eq:uniform_solution}).

The action (\ref{eq:action_definition}), in terms of $ m $ and $\ell$, is
\begin{equation}
\label{eq:S_as_function_of_kappa_and_L}
S\simeq\frac{1}{2}\int_{-\ell}^{\ell}dx\,\rho_{0}^{2}\left(x\right)=\frac{16\EllipticK^{3}\left(m\right)\left[2\left(2-m\right)\EllipticE\left(m\right)-\left(1-m\right)\EllipticK\left(m\right)\right]}{3\ell^{3}},
\end{equation}
where $\EllipticE( m )$  is the complete elliptic integral of the second kind \citep{elliptic_integrals}.
As shown in Appendix \ref{Appendix:TW}, the slope of the traveling $h$-front is given by
\begin{equation}
\label{eq:V0_as_function_of_rho0_and_kappa}
V_{0}\left(x\right)=\frac{2 m \EllipticK\left( m \right)\text{sn}\left[\frac{\EllipticK\left( m \right)x}{\ell},\, m \right]\text{cn}\left[\frac{\EllipticK\left( m \right)x}{\ell},\, m \right]}{\ell\,\text{dn}\left[\frac{\EllipticK\left( m \right)x}{\ell},\, m \right]},
\end{equation}
where $\text{sn}(u,  m )$ and $\text{cn}(u,  m )$ are Jacobi elliptic functions \citep{elliptic_functions}.
Integrating Eq.~(\ref{eq:V0_as_function_of_rho0_and_kappa}) with respect to $x$, we obtain the traveling $h$-profile
\begin{equation}
\label{eq:h0_as_function_of_rho0_and_kappa}
h_{0}\left(x\right)=-2\ln\left\{ \text{dn}\left[\frac{\EllipticK\left( m \right)x}{\ell},\, m \right]\right\} ,
\end{equation}
where we have omitted an additive constant by arbitrarily setting $h_{0}\left(0\right)=0$.

\begin{figure}[ht]
\includegraphics[width=0.55\textwidth,clip=]{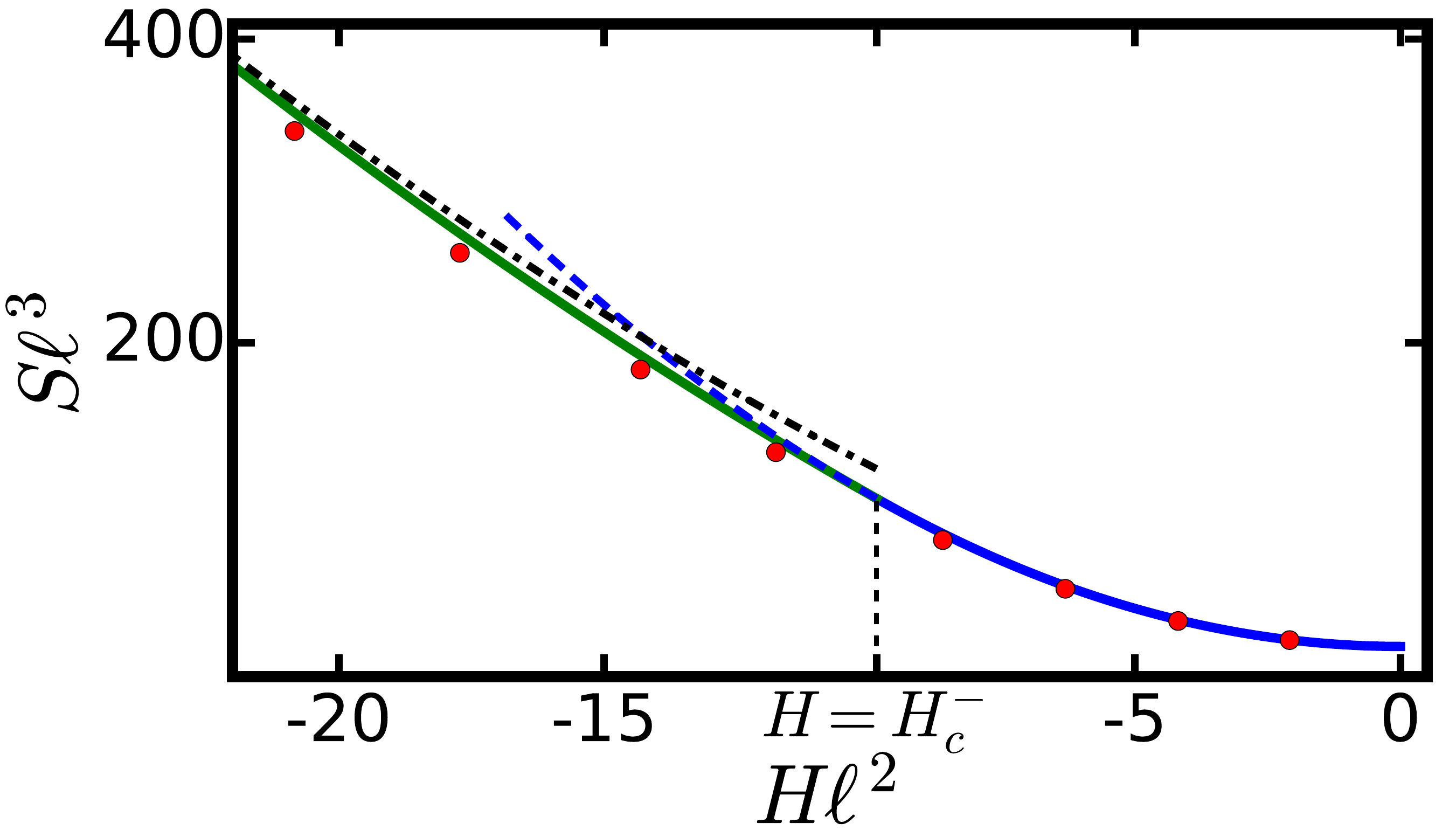}
\caption{Solid line: $S\ell^{3}$ as a function of $H\ell^{2}$ for the traveling $h$-front solution for $\lambda H > 0$ , as described by the scaling function $f$, see Eq.~(\ref{eq:scaling_function_def_TW}), obtained from  Eqs.~(\ref{eq:c_as_function_of_kappa}) and~(\ref{eq:S_as_function_of_kappa_and_L}) for supercritical $H$, and from Eq.~(\ref{eq:action_uniform_solution}) for subcritical $H$.
At $H=H_c^-$ there is a jump in the second derivative of $f$.
Also shown are the large-$|H|$ asymptotic~(\ref{eq:MKV_TW_action}) of the traveling $h$-front result (dot-dashed), and the action~(\ref{eq:action_uniform_solution}) for the uniform solution for supercritical values of $H$ (dashed).
The circles describe results obtained from a numerical solution to the OFM equations~(\ref{eq:OFM_h}) and~(\ref{eq:OFM_rho}) for $\ell=0.5$. We used
the Chernykh-Stepanov back-and-forth iteration algorithm \citep{Chernykh}.}
\label{fig:S_vs_H_TW}
\end{figure}

As one can see from Eqs.~(\ref{eq:c_as_function_of_kappa}) and~(\ref{eq:S_as_function_of_kappa_and_L}),
the action of the traveling $h$-front solution exhibits the following scaling behavior:
\begin{equation}
\label{eq:scaling_function_def_TW}
S\left(H,\ell\right)=\frac{f\left(H\ell^{2}\right)}{\ell^{3}}.
\end{equation}
For subcritical $H$, Eq.~(\ref{eq:scaling_function_def_TW}) is still valid, with $f(\alpha)=\alpha^2$, by virtue of Eq.~(\ref{eq:action_uniform_solution}).
Plugging Eq.~(\ref{eq:scaling_function_def_TW}) into Eq.~(\ref{eq:action_scaling}), we obtain the scaling form of the probability distribution in the traveling $h$-front regime, in agreement with our prediction~(\ref{eq:handwaving_positive_Hscaling}). The scaling function $f(\alpha)$ is plotted in Fig.~\ref{fig:S_vs_H_TW}.

We now present asymptotics of the scaling function $f$ in the supercritical region, obtained
from Eqs.~(\ref{eq:c_as_function_of_kappa}) and~(\ref{eq:S_as_function_of_kappa_and_L}).
The weakly-supercritical and strongly-supercritical regimes are obtained from the $ m  \to0^+$ and $ m  \to1^-$ (respectively) asymptotics of the elliptic integrals $\EllipticK( m )$ and $\EllipticE( m )$.
In the weakly-supercritical regime, $H_{c}-H\ll\left|H_{c}\right|$,  Eqs.~(\ref{eq:c_as_function_of_kappa}) and~(\ref{eq:S_as_function_of_kappa_and_L}) become
\begin{eqnarray}
\label{eq:weakly_supercritical_Hl_squared}
-H\ell^{2}&=&\pi^{2}+\frac{3\pi^{2} m ^{2}}{32}+\frac{3\pi^{2} m ^{3}}{32}+\frac{699\pi^{2} m ^{4}}{8192}+\dots,\\
S\ell^{3}&=&\pi^{4}+\frac{3\pi^{4} m ^{2}}{16}+\frac{3\pi^{4} m ^{3}}{16}+\frac{711\pi^{4} m ^{4}}{4096}+\dots,
\end{eqnarray}
which, using Eq.~(\ref{eq:Hcminus_def}), leads to
\begin{equation}
S=S\left(H_{c}^{-}\right)+\frac{2\pi^{2}\left(H_{c}^{-}-H\right)}{\ell}+\frac{\left(H_{c}^{-}-H\right)^{2}\ell}{3}+\dots.
\end{equation}
In the strongly supercritical regime,
\begin{equation}
\label{eq:strongly_nonlinear_limit}
\left|H\right| \! \ell^{2}\gg1,
\end{equation}
Eqs.~(\ref{eq:c_as_function_of_kappa}) and~(\ref{eq:S_as_function_of_kappa_and_L}) reduce to
\begin{equation}
\label{eq:strongly_supercritical_Hl_squared}
H\ell^{2} \simeq -\frac{1}{2}\,\ln^{2}\left(\frac{1- m }{16}\right)\quad \mbox{and}\quad S\ell^{3} \simeq \frac{4}{3}\,\ln^{3}\left(\frac{1- m }{16}\right),
\end{equation}
yielding
\begin{equation}
\label{eq:MKV_TW_action}
S\simeq\frac{8\sqrt{2}}{3}\left|H\right|^{3/2},
\end{equation}
or, in the dimensional variables,
\begin{equation}
-\ln\mathcal{P}\left(H,L,T\right)\simeq\frac{8\sqrt{2}\nu\left|H\right|^{3/2}}{3D\left|\lambda\right|^{1/2}T^{1/2}}.
\end{equation}
This tail is universal: it coincides with the slower-decaying tails of the GOE and GUE Tracy-Widom distributions \citep{CLD, Sasamoto2010,Calabrese2010,Dotsenko2010,Amir2011}
and holds for a broad class of deterministic initial conditions \citep{KMSparabola}. The absence of finite-$L$ effects
in this regime is explained by the fact that the width of the soliton of $\rho\left(x,t\right)=\rho_{0}\left(x\right)$ scales as $\left|H\right|^{-1/2}$ \citep{KK2007,MKV}, see below. At $\left|H\right| \! \ell^{2}\gg1$ the soliton is strongly localized within the ring and does not feel the system size. The optimal height profile $h\left(x,t\right)$ feels the system size only via a small correction, see below. The action $S$ is unaffected up to an exponentially small correction.

Altogether, the function $f(\alpha)$ from Eq.~(\ref{eq:scaling_function_def_TW}) has the following asymptotic behavior:
\begin{equation}
f\left(\alpha\right)=\begin{cases}
\alpha^{2}, & -\pi^{2}=\alpha_{c}<\alpha<0,\\
\alpha_{c}^{2}+2\pi^{2}\left(\alpha_{c}-\alpha\right)+\frac{\left(\alpha_{c}-\alpha\right)^{2}}{3}+\dots, & \alpha_{c}-\alpha\ll\left|\alpha_{c}\right|,\\
\frac{8\sqrt{2}\,\left|\alpha\right|^{3/2}}{3}+\dots, & \alpha<0,\quad\left|\alpha\right|\gg1.
\end{cases}
\end{equation}
Notable is a discontinuity of the second derivative $f''\left(\alpha\right)$ at $\alpha = \alpha_c$, implying  a second-order dynamical phase transition in the {traveling $h$-front action $S$ at $\left|H\right| = \left|H_{c}^{-}\right|$. As we will argue shortly, a true phase transition in the complete OFM formulation appears only in the limit $\ell \to 0$.

In the weakly supercritical limit, the leading terms in the
Fourier series expansion of $\text{dn}^2$ in Eq.~(\ref{eq:rho0_negative_tail_exact}) yield
\begin{equation}
\rho_{0}\left(x\right)=-\frac{\pi^{2}}{\ell^{2}}\left[1+\zeta-\zeta^{2}+\left(\sqrt{8\zeta}+\frac{\zeta^{3/2}}{4\sqrt{2}}\right)\cos\left(\frac{\pi x}{\ell}\right)+\left(2\zeta+\dots\right)\cos\left(\frac{2\pi x}{\ell}\right)+\dots\right],
\end{equation}
where $\zeta=\left(H_{c}^{-}-H\right)\ell^{2}/\left(3\pi^{2}\right)$, and we have omitted all terms whose contribution to the action is of a higher order in $\zeta$  than quadratic.
The optimal profile~(\ref{eq:h0_as_function_of_rho0_and_kappa}) in this regime exhibits a small modulation:
\begin{equation}
\label{eq:h0_weakly_nonlinear}
h_{0}\left(x\right) \simeq \sqrt{8\zeta}\left[1-\cos\left(\frac{\pi nx}{\ell}\right)\right].
\end{equation}
In the strongly-supercritical regime~(\ref{eq:strongly_nonlinear_limit}) Eq.~(\ref{eq:rho0_negative_tail_exact}) yields the soliton of Refs. \cite{KK2007,MKV}, as to be expected:
\begin{equation}
\label{eq:rho_0_strongly_nonlinear}
\rho_{0}\left(x\right)\simeq-2\left|H\right|\text{sech}^{2}\left(\sqrt{\frac{\left|H\right|}{2}}\:x\right),
\end{equation}
The optimal $h$-profile becomes
\begin{equation}
\label{eq:optimal_profile_sol_TW}
h\left(x,t\right)=h_{0}\left(x\right)-ct\simeq\begin{cases}
\;2\ln\cosh\left(\sqrt{\frac{\left|H\right|}{2}}\:x\right)-ct, & \ell-\left|x\right|\gg\frac{1}{\sqrt{H}}\,,\\
-2\ln\cosh\left[\sqrt{\frac{\left|H\right|}{2}}\:\left(\ell-\left|x\right|\right)\right]+\sqrt{2\left|H\right|}\,\ell-4\ln2-ct, & \;\left|x\right|\gg\frac{1}{\sqrt{H}}\;.
\end{cases}
\end{equation}
As to be expected, this result also coincides with the corresponding result of Ref. \citep{MKV}, except in a narrow boundary layer around $x=\pm \ell$, where $h(x,t)$ readily adapts to the periodic boundary conditions.

\subsection{Numerics}

The traveling $h$-front solution, presented in the previous subsection, solves the OFM equations exactly, but it does not obey all of the temporal boundary conditions
of the full problem. Therefore, in general, it is not optimal. To identify the regime of parameters $(H,\ell)$ where the traveling $h$-front solution is a good approximation to the optimal solution, we solved the full problem numerically using the Chernykh-Stepanov back-and-forth iteration algorithm \citep{Chernykh}.
For $\ell \ll 1$ the numerical solutions are in good agreement with the traveling $h$-front solution for all $H<0$.
In particular, the numerics support our prediction of the second-order dynamical phase transition at $H=H_c^-$~(\ref{eq:Hcminus_def}) in the limit $\ell\to0$ \footnote{A true singularity of $S$ at $H=H_c^-$, and the ensuing phase transition, only occur in the limit $\ell\to0$, where the traveling $h$-front solution becomes exact. For nonzero $\ell$, there is a temporal boundary layer near $t=1$ where $\rho$ rapidly adapts to the boundary condition~(\ref{eq:delta_initial_cond_rho}), and, for supercritical $H$, an additional boundary layer near $t=0$, where $h$ adapts to the condition~(\ref{eq:flat_initial_cond}). These boundary layers smoothen out the transition.}. See Fig.~\ref{fig:S_vs_H_TW}, where the numerically-computed action for $\ell=0.5$ is compared with the prediction of Eq.~(\ref{eq:scaling_function_def_TW}), and Fig. \ref{fig:optimal_trajetory_TW2}, where $h(x,t)$ is plotted for $\ell=0.5$ and $\Lambda=-48$.

\begin{figure}[ht]
\includegraphics[width=0.42\textwidth,clip=]{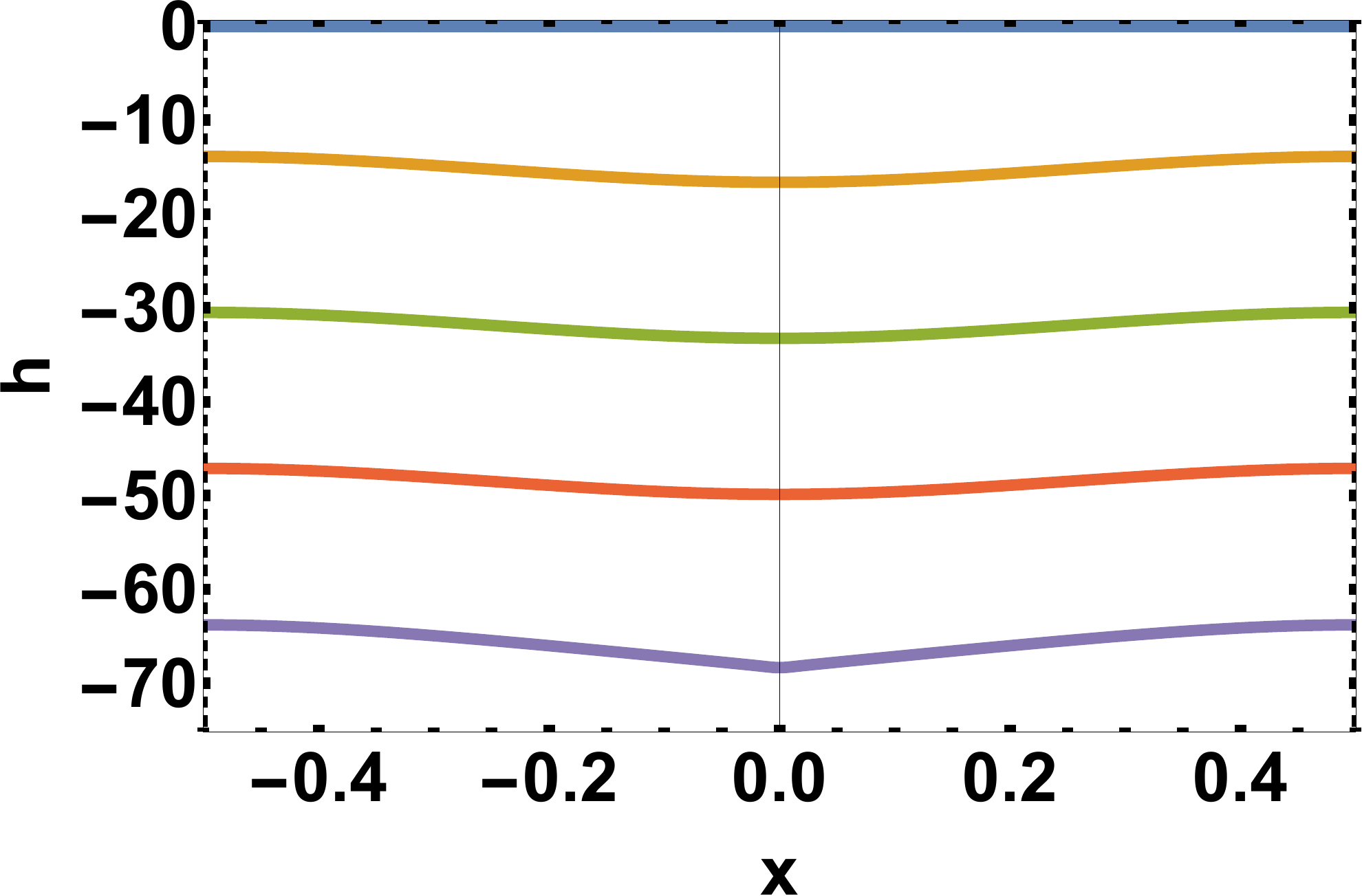}
\llap{\parbox[b]{4.7in}{{\Large (a)}\\\rule{0ex}{1.61in}}}
\includegraphics[width=0.4\textwidth,clip=]{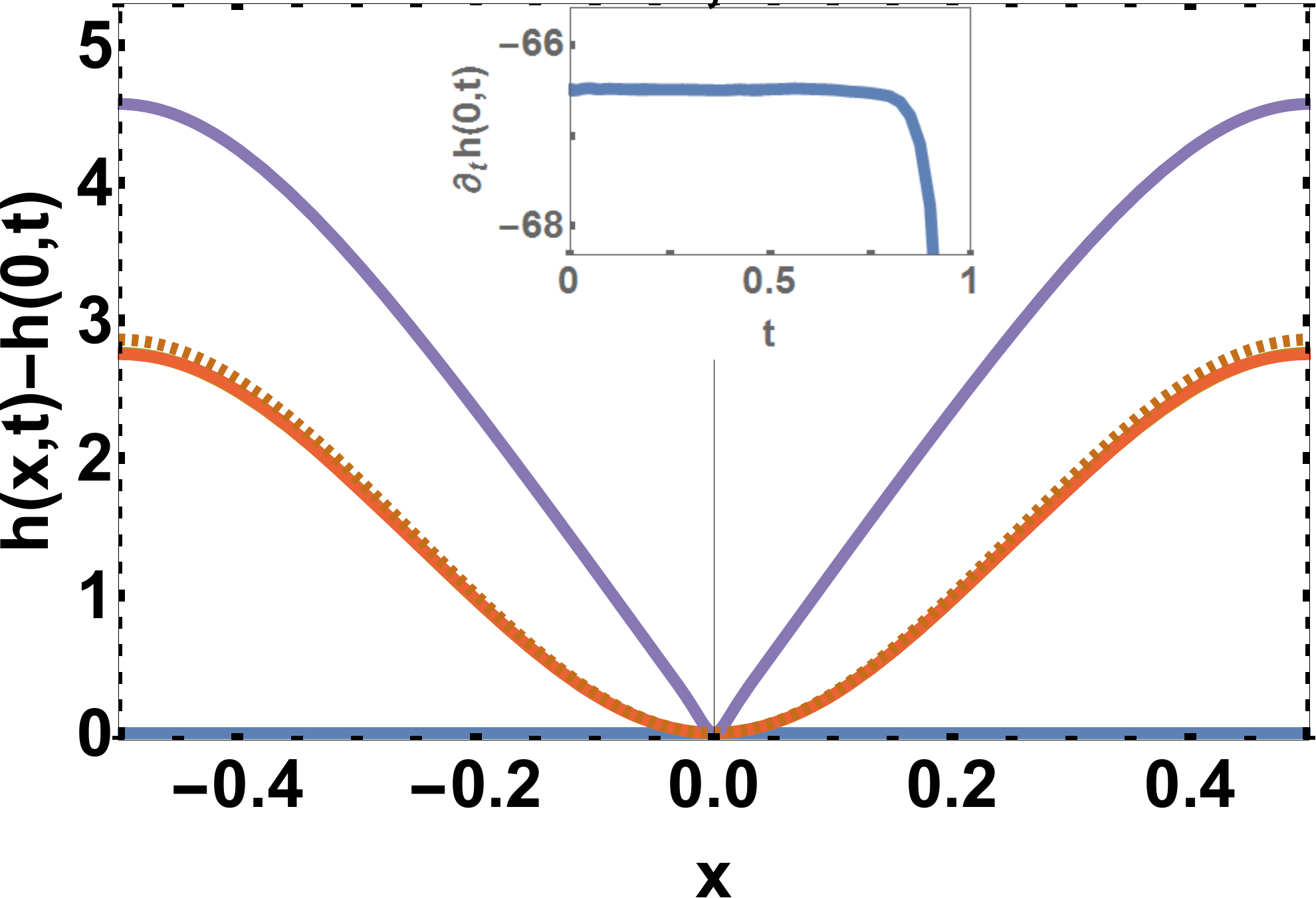}
\llap{\parbox[b]{4.95in}{{\Large (b)}\\\rule{0ex}{1.35in}}}
\includegraphics[width=0.4\textwidth,clip=]{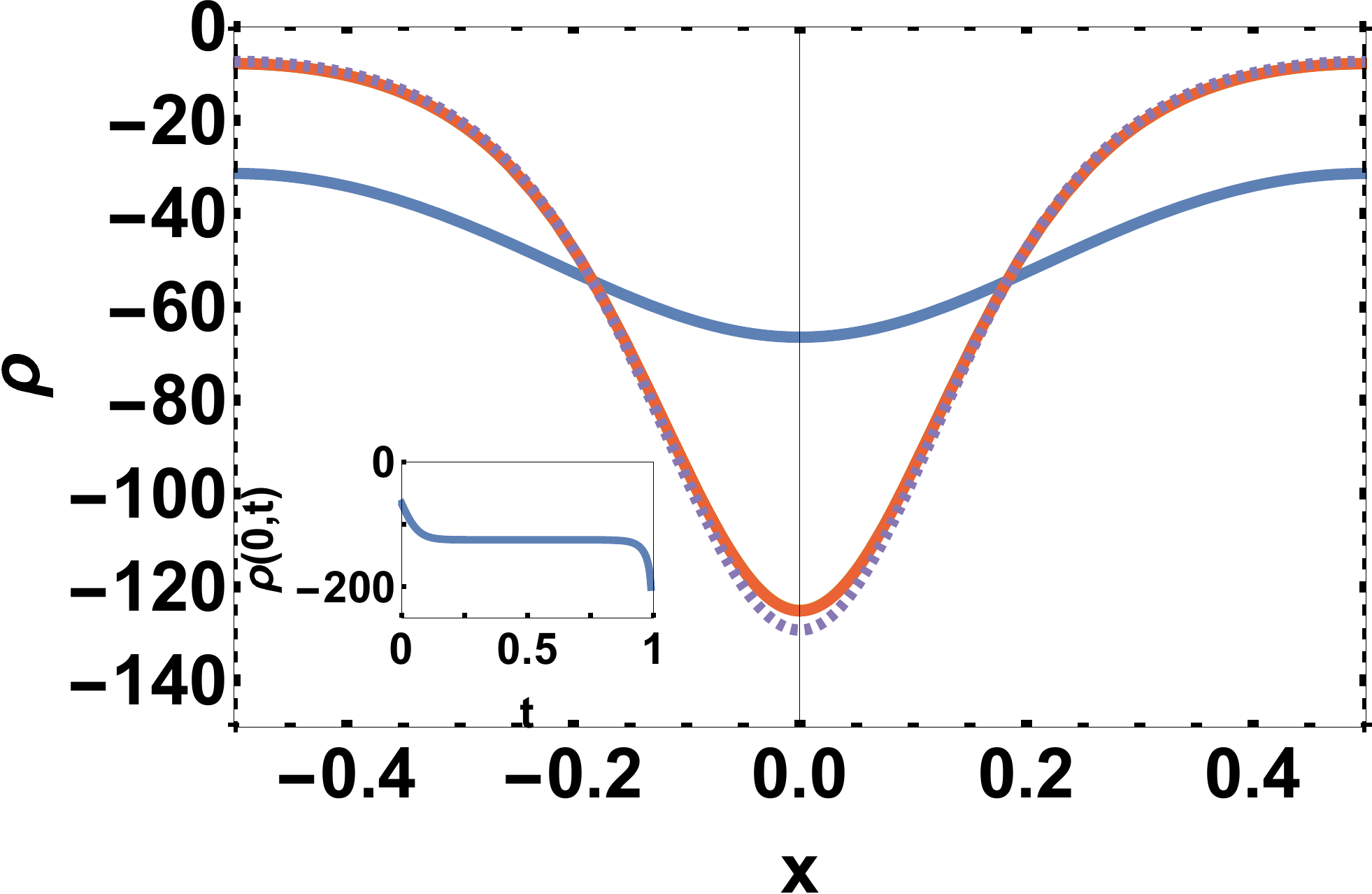}
\llap{\parbox[b]{3.4in}{{\Large (c)}\\\rule{0ex}{1.5in}}}
\caption{The optimal interface history for large $\lambda H > 0$. Parameters are $\ell=0.5$ and $\Lambda=-48$. (a) $h$ vs. $x$ at times $t=0, 0.25T, 0.5T, 0.75T$ and $T$ (from top to bottom). (b) $h(x,t) - h(0,t)$ vs. $x$ for the same times as in (a) (from bottom to top). There is a collapse of the curves for all times except $t=0$ and $t=T$. Inset: $\partial_th(0,t)$ vs.~$t$. (c) $\rho$ vs.~$x$ at times $t=0, 0.25T, 0.5T$ and $0.75T$. Inset: $\rho(x=0,t)$. The analytical (dashed) and numerical (solid) curves are almost indistinguishable, except at $t=T$ for (b) and at $t=0$ for (c). The velocity of the traveling $h$-front, $c=66.5$, is within $3\%$ of $|H|$, where $H=h(0,T)=-68.3$ is the final interface height at the origin. The action, computed numerically, is $S\simeq2073$, while the analytic prediction from Eq.~(\ref{eq:S_as_function_of_kappa_and_L}) is $S\simeq2051$.
}
\label{fig:optimal_trajetory_TW2}
\end{figure}

\begin{figure}[ht]
\includegraphics[width=0.4\textwidth,clip=]{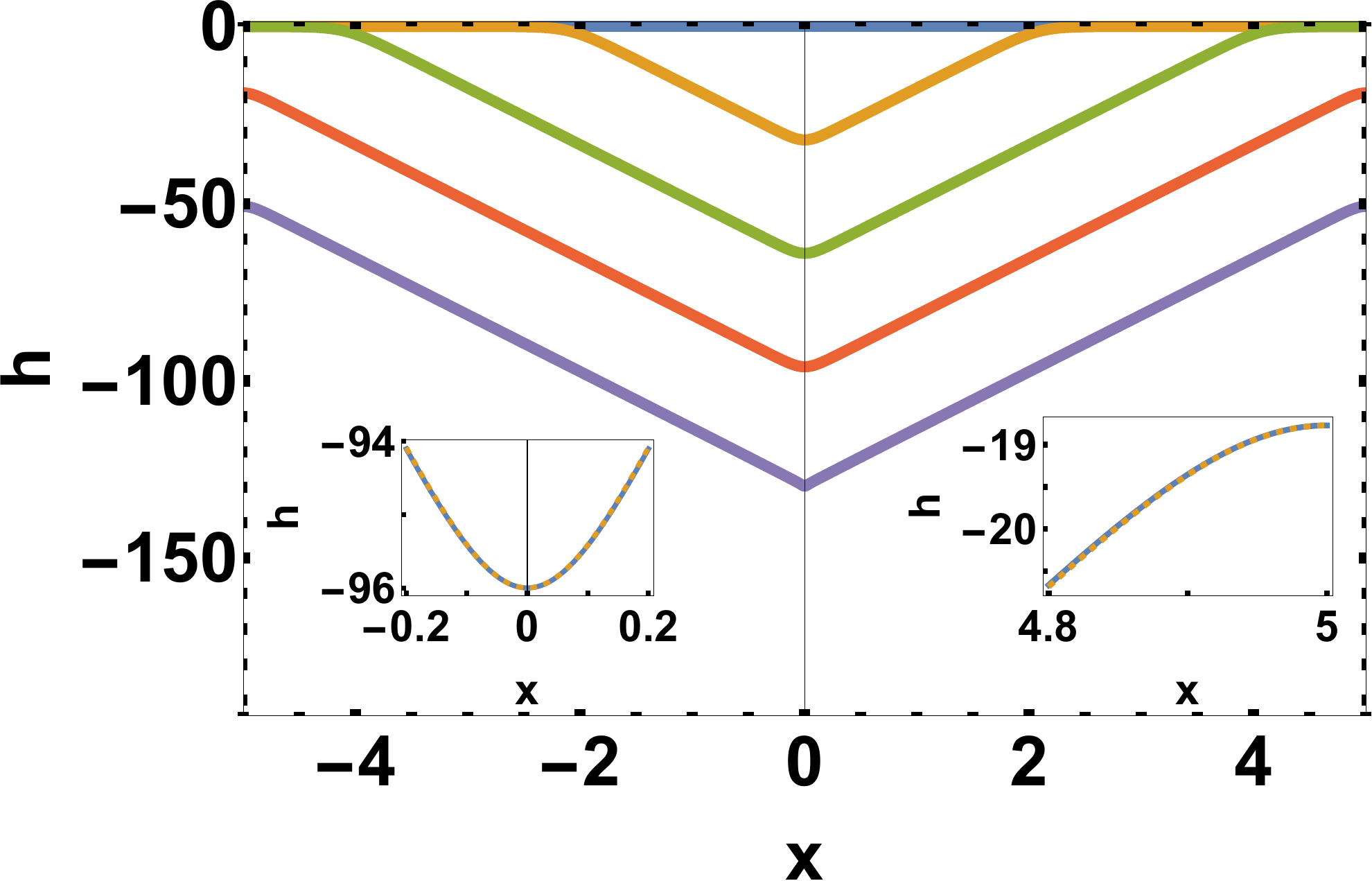}
\llap{\parbox[b]{5.2in}{{\Large (a)}\\\rule{0ex}{1.5in}}}
\includegraphics[width=0.4\textwidth,clip=]{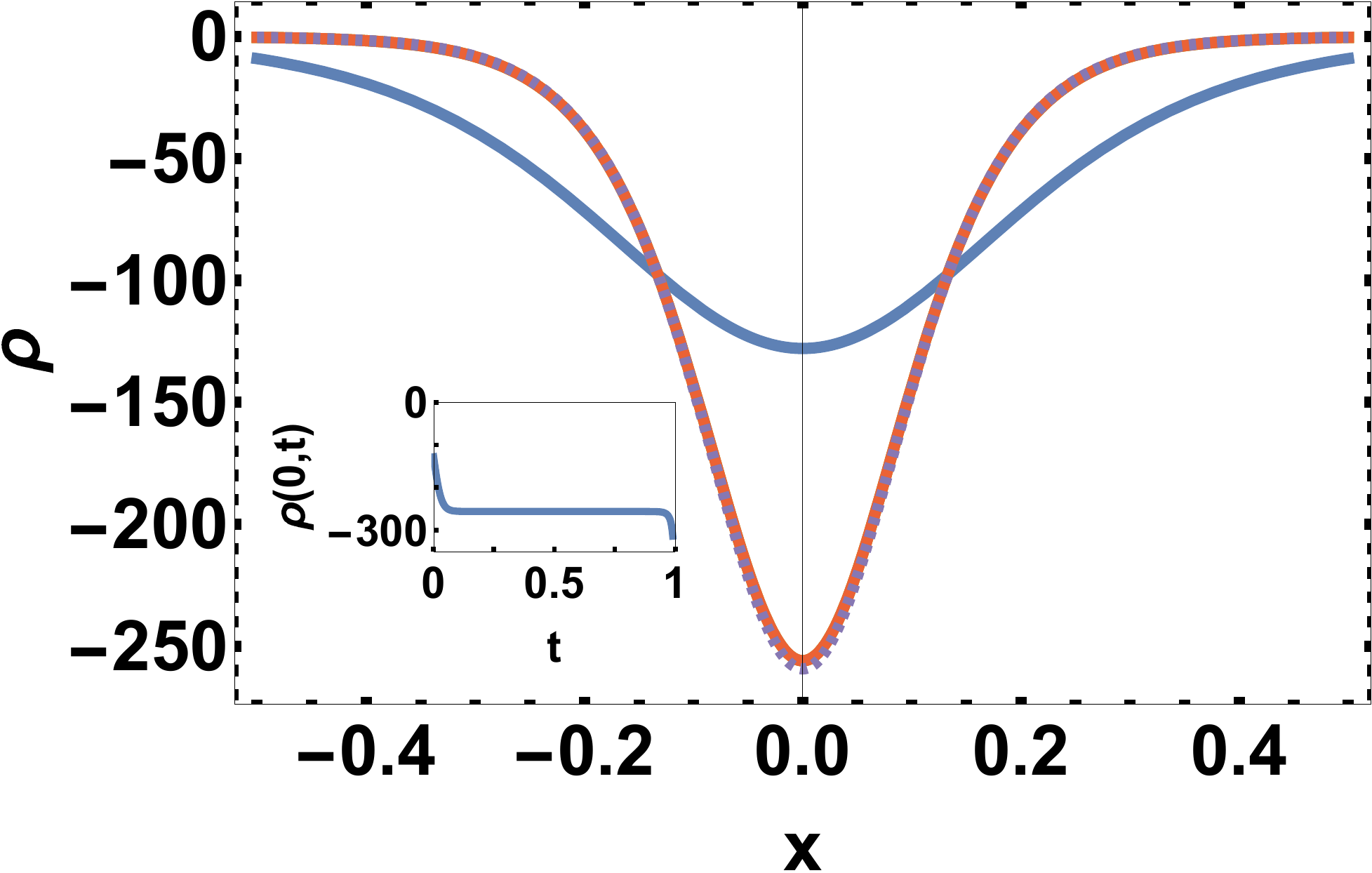}
\llap{\parbox[b]{4.5in}{{\Large (b)}\\\rule{0ex}{1.35in}}}
\caption{The optimal interface history for $\lambda H \to +\infty$ and $\ell \gg 1$. Parameters are $\ell=5$ and $\Lambda=-64$. (a) $h$ vs. $x$ at times $t=0, 0.25T, 0.5T, 0.75T, T$. Insets: the boundary layers at $x=0$ and at $x=\ell$ at $t=0.75T$. (b) $\rho$ vs. $x$ at times $t=0, 0.25T, 0.5T, 0.75T$. Inset: $\rho(x=0,t)$. The analytical (solid) and numerical (dashed) curves are indistinguishable, except at $t=0$ for (b).}
\label{fig:optimal_trajetory_TW}
\end{figure}

For $\ell \gtrsim 1$ the numerical solution shows a traveling $h$-front only in the strongly supercritical regime~(\ref{eq:strongly_nonlinear_limit}). For $1\ll\left|H\right|<2\ell^{2}$, the solution is in good agreement with the traveling $h$-front for an infinite system, found in Refs. \cite{KK2007,MKV}, and there are no finite-size effects. $S$ and $\rho_0$ are in good agreement with Eqs.~(\ref{eq:MKV_TW_action}) and~(\ref{eq:rho_0_strongly_nonlinear}), respectively.
For $\left|H\right| > 2\ell^{2}\gg1$, the optimal interface $h\left(x,t\right)$ behaves as the traveling $h$-front solution for an infinite system \citep{KK2007,MKV} until (rescaled) time $t=\sqrt{2}\,\ell/\sqrt{\left|H\right|} < 1$, when $h$ starts to feel finite-$\ell$ effects. From this time on, $h\left(x,t\right)$ takes the form~(\ref{eq:optimal_profile_sol_TW}). This is shown in Fig. \ref{fig:optimal_trajetory_TW} for $\ell=5$ and $H\simeq-130$. In Fig.~\ref{fig:phase_diagram} we show a phase diagram of the system in the $(L/\sqrt{T}, H)$ plane. The tail $\lambda H \to -\infty$ is dealt with in the next section.

\begin{figure}[ht]
\includegraphics[width=0.4\textwidth,clip=]{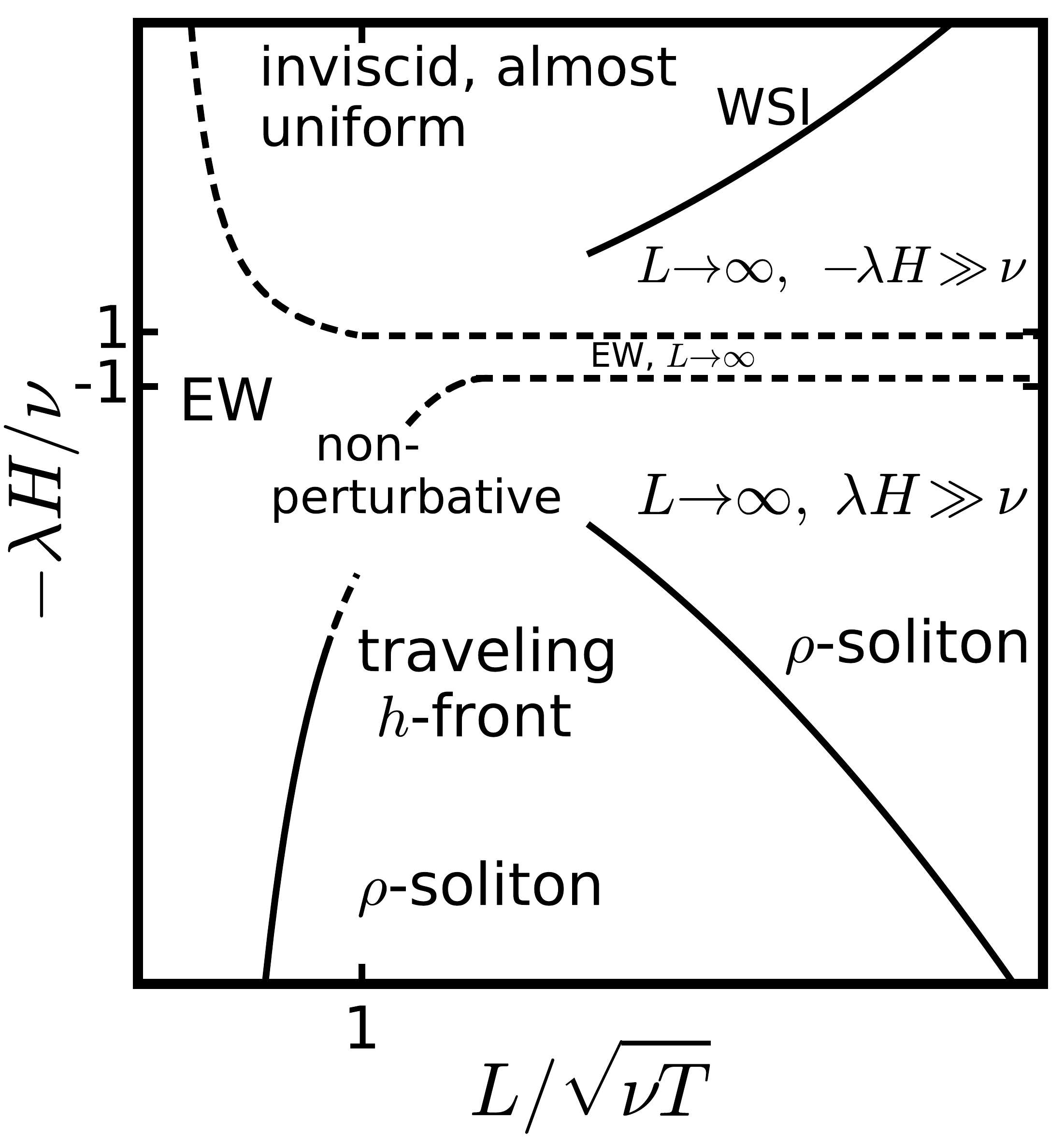}
\caption{Phase diagram in the $(L/\sqrt{T}, H)$ plane. The solid lines describe sharp transitions, the dashed lines denote gradual crossovers. In the ``non-perturbative" region -- where $\ell$ and the rescaled $H$ are both of order unity -- and in the vicinity of the dashed lines, analytic solutions to the OFM equations are unavailable. In all other regions we found perturbative solutions. In the regions marked $L\to\infty$ the system behaves as if it were infinite.
In the Edwards-Wilkinson (EW) regime, the height fluctuations are Gaussian, see Sec. \ref{sec:typical}. For large  $\lambda H > 0$, the optimal trajectory is a traveling $h$-front, see Sec. \ref{sec:negative_tail}, and for $\lambda H \to - \infty$, the dynamics are approximately inviscid, and the optimal trajectory is nearly uniform, see Sec.~\ref{sec:positive_tail_perturbation_theory}.
The weakly-supercritical inviscid (WSI) region is discussed in Sec. \ref{section:small_hole}.}
\label{fig:phase_diagram}
\end{figure}

\section{Large negative $\lambda H$}
\label{sec:positive_tail}

The regime of large negative $\lambda H$ has a very different nature. Here one can neglect, in the leading order, the
diffusion terms in Eqs.~(\ref{eq:OFM_h}) and~(\ref{eq:OFM_rho}) \citep{KK2007,KK2009,MKV,KMSparabola, Janas2016} thus arriving at an inviscid hydrodynamic formulation. We begin this section with this formulation, present the `inviscid scaling' of the action and apply the hodograph transformation. In Sec.~\ref{sec:positive_tail_perturbation_theory} we solve this problem in the limit $\lambda H \to - \infty$ (for fixed $\ell=L/\sqrt{\nu T}$). In Sec.~\ref{section:small_hole} we show that, in the large-$\ell$ limit, the negative $\lambda H$ tail exhibits a fractional-order phase transition at an $\ell$-dependent value $H_c^+$, where finite-size effects become important, and evaluate the action in the weakly supercritical limit.

\subsection{Hydrodynamic analogy}
\label{sec:hydrodynamic_analogy}

For large negative $\lambda H$, one can drop the
diffusion terms in Eqs.~(\ref{eq:OFM_h}) and~(\ref{eq:OFM_rho}) \citep{KK2007,KK2009,MKV,KMSparabola, Janas2016} and arrive at
\begin{eqnarray}
\label{eq:OFM_V_positive}
  \partial_t V +V \partial_x V &=&\partial_x \rho,\\
\label{eq:OFM_rho_positive}
 \partial_t \rho +\partial_x (\rho V)&=& 0,
\end{eqnarray}
where $V\left(x,t\right)=\partial_{x}h\left(x,t\right)$.
These equations describe an inviscid hydrodynamic flow of an effective gas with density $\rho$, velocity $V$ and a \emph{negative} gas pressure. They describe a universal long-wavelength hydrodynamic instability
which appears in many applications \cite{Trubnikov1987}.  Remarkably, the same equations also describe a far tail
of the probability distribution of time-integrated current through a lattice site in the Simple Symmetric Exclusion Process (SSEP) on an infinite line \cite{MS2014,VMS2014}.
In our problem Eqs.~(\ref{eq:OFM_V_positive}) and (\ref{eq:OFM_rho_positive}) should be solved with the initial condition $V\left(x,0\right)=0$, periodic spatial boundary conditions, and the temporal boundary condition~(\ref{eq:delta_initial_cond_rho}).

In an infinite system, the inviscid solution  has a compact support in $x$ \cite{KK2009,MKV}. It describes compression of a ``cloud" of the effective gas, which is initially at  $|x| \leq 2\sqrt{2H}/\pi$. This compression culminates at $t=1$ when all of the gas collapses
into the point $x=0$ in compliance with Eq.~(\ref{eq:delta_initial_cond_rho}).
Therefore, for a fixed $\ell$, there is a critical value of $H$,
\begin{equation}
\label{eq:Hcplus_def}
H_{c}^{+}=\frac{\pi^{2}\ell^{2}}{8},
\end{equation}
so that for $H>H_{c}^{+}$  the infinite-system-size solution~\cite{KK2009,MKV} does not fit into the ring. This implies important finite-size effects which will be our focus here.

The rescaling transformation \citep{MKV}
$x/\sqrt{H}\to x$, $V/\sqrt{H}\to V$, $\rho/H\to\rho$ and $h/H\to h$ keeps Eqs.~(\ref{eq:OFM_V_positive}) and~(\ref{eq:OFM_rho_positive}) invariant, but leaves only one parameter in the problem -- the rescaled system size $\ell/\sqrt{H}$. The rescaled action  (\ref{eq:action_definition}), therefore, exhibits  ``inviscid scaling":
\begin{equation}
\label{eq:action_scaling_positive_tail}
S\left(H,\ell\right)=H^{5/2}s\left(\frac{\ell}{\sqrt{H}}\right),
\end{equation}
so that, in the dimensional variables, the large deviation function of height is dominated
by the KPZ nonlinearity and independent of $\nu$, verifying our prediction~(\ref{eq:handwaving_inviscid}). As we find below in this Section, the scaling function $s(\alpha)$ behaves as follows:
\begin{equation}
\label{eq:scaling_function_s_asymptotics}
s\left(\alpha\right)=\begin{cases}
\frac{8\sqrt{2}}{15\pi}, & \alpha>\frac{2\sqrt{2}}{\pi},\\
\frac{8\sqrt{2}}{15\pi}-c_{2}\left(\frac{2\sqrt{2}}{\pi}-\alpha\right)^{5/2}+\dots, & 0<\frac{2\sqrt{2}}{\pi}-\alpha\ll1,\\
\alpha-\frac{16\alpha^{2}}{3\pi}+\dots, & 0<\alpha\ll1,
\end{cases}
\end{equation}
with $c_{2}\simeq0.50$. For $\ell\gg1$  and $1\ll H<H_{c}^{+}$ the solution coincides with that for an infinite system, so the first line of Eq.~(\ref{eq:scaling_function_s_asymptotics}) describes the $L$-independent result~\cite{KK2009,MKV}
\begin{equation}
\label{eq:positive_tail_no_finite_size_effects}
-\ln\mathcal{P}\simeq\frac{8\sqrt{2\left|\lambda\right|}}{15\pi D}\frac{H^{5/2}}{T^{1/2}}.
\end{equation}

To solve the problem for $H>H_{c}^{+}$, we will use the hodograph transformation \citep{Courant1948,Trubnikov1987}, where $\rho$ and $V$ are treated as the independent variables, and $t$ and $x$ as the dependent
ones. The hodograph transformation transforms the nonlinear Eqs.~(\ref{eq:OFM_V_positive}) and~(\ref{eq:OFM_rho_positive}) into linear equations
\begin{eqnarray}
\label{eq:OFM_dx_dV}
\partial_{V}x&=&V\partial_{V}t-\rho\partial_{\rho}t,\\
\label{eq:OFM_dx_drho}
\partial_{\rho}x&=&\partial_{V}t+V\partial_{\rho}t,
\end{eqnarray}
for $x\left(\rho,V\right)$ and $t\left(\rho,V\right)$. The additional change of variables $z=V/2$ and $r=\sqrt{\rho}$ brings Eqs.~(\ref{eq:OFM_dx_dV}) and~(\ref{eq:OFM_dx_drho}) to the form:
\begin{eqnarray}
\label{eq:OFM_dx_dz}
\partial_{z}x&=&2z\partial_{z}t-r\partial_{r}t, \\
\label{eq:OFM_dx_dr}
\partial_{r}x&=&r\partial_{z}t+2z\partial_{r}t.
\end{eqnarray}
These two first-order equations yield a single second-order elliptic equation for $t\left(r,z\right)$ \cite{Trubnikov1987},
\begin{equation}
\label{eq:Laplace_like_t_z_r}
\partial_{z}^{2}t+\partial_{r}^{2}t+\frac{3\partial_{r}t}{r}=0,
\end{equation}
which should be solved in the half-plane $\left|z\right|<\infty$, $0\le r<\infty$. What are the boundary conditions? At $t=0$ $V\left(x,t=0\right)=0$, while $\rho\left(x,t=0\right)$ varies between two (a priori unknown) values $0<\rho_{1}<\rho_{2}$.
These conditions yield the Dirichlet condition
\begin{equation}
\label{eq:t_Dirichlet_condition}
t\left(r,z=0\right)=0, \qquad r_1 \le r \le r_2,
\end{equation}
where $r_{i}=\sqrt{\rho_{i}}$, $i=1,2$.
Further, the boundary condition~(\ref{eq:delta_initial_cond_rho}) translates into
\begin{equation}
\label{eq:t_boundary_condition_at_infinity}
t\left(r\to\infty,z\right)=t\left(r,\left|z\right|\to\infty\right)=1.
\end{equation}
Finally, $t\left(r,z\right)$ must be regular at $r=0$, to enable matching of the hodograph solution with the solution of the Hopf equation $\partial_tV+V\partial_xV=0$ in the regions where $\rho=0$ \cite{MKV}.

\subsection{Electrostatic analogy}
\label{sec:electrostatic}

Following Ref. \cite{Trubnikov1987}, we introduce an auxiliary angle $0\le\phi<2\pi$ and define a new dependent variable
\begin{equation}
\label{eq:Psi_def}
\Psi\left(r,z,\phi\right)=\tbar\left(r,z\right)r\cos\phi,
\end{equation}
where $\tbar\left(r,z\right) = 1-t\left(r,z\right)$.
It is easy to see from Eqs.~(\ref{eq:Laplace_like_t_z_r}) and~(\ref{eq:Psi_def}) that $\Psi\left(r,z,\phi\right)$ obeys the 3-dimensional Laplace equation
\begin{equation}
\label{eq:Psi_laplace}
\nabla^{2}\Psi=\frac{1}{r}\frac{\partial}{\partial r}\left(r\frac{\partial\Psi}{\partial r}\right)+\frac{1}{r^{2}}\frac{\partial^{2}\Psi}{\partial\phi^{2}}+\frac{\partial^{2}\Psi}{\partial z^{2}}=0,
\end{equation}
where $\left(r,z,\phi\right)$ are interpreted as cylindrical coordinates.
By virtue of Eq.~(\ref{eq:t_Dirichlet_condition}), the harmonic function $\Psi\left(r,z,\phi\right)$ should obey a Dirichlet boundary condition on an \emph{annulus} of inner radius $r_1$ and outer radius $r_2$ (see Fig. \ref{fig:annulus}):
\begin{equation}
\label{eq:Psi_Dirichlet_condition}
\Psi\left(r_{1}\le r\le r_{2},z=0,\phi\right)=-r\cos\phi.
\end{equation}
The boundary condition~(\ref{eq:t_boundary_condition_at_infinity}) at infinity becomes
\begin{equation}
\label{eq:Psi_boundary_condition_at_infinity}
\Psi\left(r\text{ or }\left|z\right|\to\infty,\phi\right)=0.
\end{equation}
The regular behavior of $t$ as $r\to0$ implies that the Laplace equation~(\ref{eq:Psi_laplace}) is satisfied on the $z$ axis as well, so $\Psi$ has no singularities there.

\begin{figure}[ht]
\includegraphics[width=0.42\textwidth,clip=]{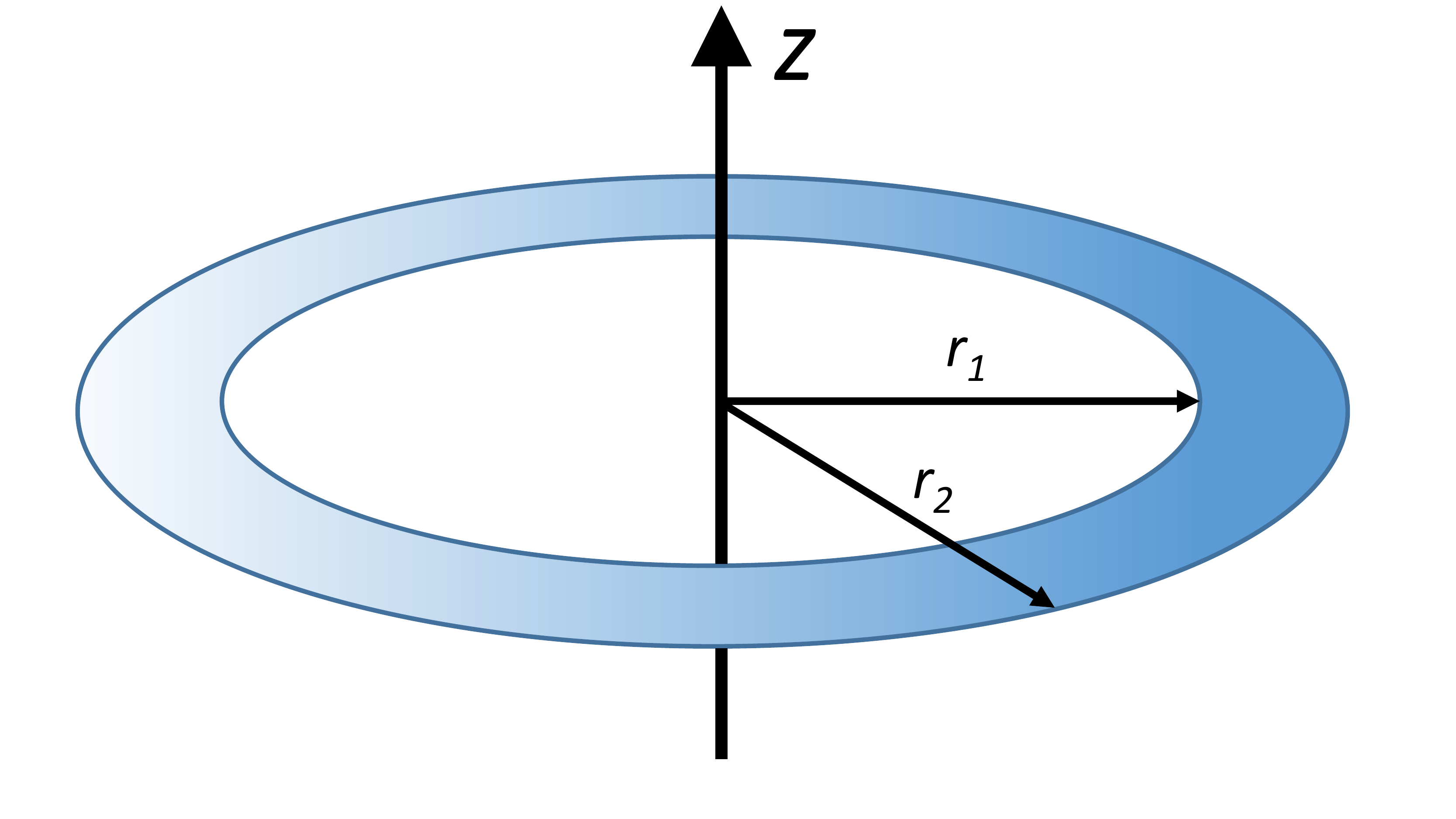}
\caption{The annulus on which the Dirichlet boundary condition~(\ref{eq:Psi_Dirichlet_condition}) is satisfied.}
\label{fig:annulus}
\end{figure}

\subsection{Effective charge density}

We interpret $\Psi\left(r,z,\phi\right)$ as an electric potential. One way of calculating it is to first determine the electric charge density on the annulus
$r_1 \le r \le r_2,\; z = 0$ so that the Dirichlet boundary condition~(\ref{eq:Psi_Dirichlet_condition}) is satisfied on both sides of the annulus. The zero boundary condition at infinity~(\ref{eq:Psi_boundary_condition_at_infinity}) will be satisfied automatically. The charge density is of the form
\begin{equation}
\label{eq:wdef}
w\left(r,z,\phi\right)=-\delta\left(z\right)g\left(r\right)\cos\phi,
\end{equation}
with an a priori unknown function $g\left(r\right) \ge 0$ which has compact support $r_1 \le r \le r_2$. $\Psi$ satisfies the Poisson equation
\begin{equation}
\label{eq:Psi_poisson}
\nabla^{2}\Psi=-4\pi w
\end{equation}
which reduces to the Laplace equation~(\ref{eq:Psi_laplace}) everywhere except on the annulus.
The solution to Eq.~(\ref{eq:Psi_poisson})
is
\begin{equation}
\label{eq:Psi_integral_over_w}
\Psi\left(\vect{r}\right)=\int\frac{w\left(\vect{r}'\right)d^{3}\vect{r}'}{\left|\vect{r}-\vect{r}'\right|}.
\end{equation}
In the derivations that follow we will need to know the asymptotic behavior of $\Psi$ at large distances from the annulus.  The leading-order multipole contribution there is the dipole:
\begin{equation}
\Psi\left(r,z,\phi\right)\sim\frac{r\cos\phi}{\left(r^{2}+z^{2}\right)^{3/2}},\qquad\sqrt{r^{2}+z^{2}}\gg\bar{r},
\end{equation}
where $\bar{r}=(r_{1}+r_{2})/2$ is the mean radius of the annulus. This implies
\begin{equation}
\label{eq:tbar_asymptotics_large_rz}
\tbar\left(r,z\right)\sim\frac{1}{\left(r^{2}+z^{2}\right)^{3/2}},\qquad\sqrt{r^{2}+z^{2}}\gg\bar{r}.
\end{equation}
We now return to the full problem of finding $\Psi$.
Using Eq.~(\ref{eq:wdef}), we evaluate the integral~(\ref{eq:Psi_integral_over_w}) in cylindrical coordinates:
\begin{eqnarray}
\label{eq:Psi_exact_solution_as_integral}
\Psi\left(r,z,\phi\right)&=&-\int_{r_{1}}^{r_{2}}dr'r'g\left(r'\right)\int_{0}^{2\pi}\frac{d\phi'\cos\phi'}{\sqrt{r^{2}+r'^{2}-2rr'\cos\left(\phi-\phi'\right)+z^{2}}}\nonumber\\
&=&-\cos\phi\int_{r_{1}}^{r_{2}}dr'r'g\left(r'\right)\int_{0}^{2\pi}\frac{d\theta\cos\theta}{\sqrt{r^{2}+r'^{2}-2rr'\cos\theta+z^{2}}}\nonumber\\
&=&-\cos\phi\int_{r_{1}}^{r_{2}}dr'r'g\left(r'\right)\mathcal{F}\left(r,r',z\right),
\end{eqnarray}
where
\begin{equation}
\label{eq:F_r_rprime_z_def}
\mathcal{F}\left(r,r',z\right)=\frac{2\left(r^{2}+r'^{2}+z^{2}\right) \EllipticK \left[-\frac{4rr'}{\left(r-r'\right)^{2}+z^{2}}\right]-2\left[\left(r-r'\right)^{2}+z^{2}\right] \EllipticE \left[-\frac{4rr'}{\left(r-r'\right)^{2}+z^{2}}\right]}{rr'\sqrt{\left(r-r'\right)^{2}+z^{2}}}.
\end{equation}
The Dirichlet boundary condition~(\ref{eq:Psi_Dirichlet_condition}) on the annulus leads to the integral equation
\begin{equation}
\label{eq:integral_equation}
\int_{r_{1}}^{r_{2}}dr'\,r'g\left(r'\right)\mathcal{F}\left(r,r',0\right)=r,\qquad r_{1}\le r\le r_{2},
\end{equation}
for the unknown function $g(r)$.

The particular case $r_1 = 0$ corresponds to $H < H_c^+$, where the solution behaves as in an infinite system \cite{KK2009,MKV}. One can show that in this case
\begin{equation}
g\left(r\right)=\frac{2r}{\pi^{2}\sqrt{r_{2}^{2}-r^{2}}} .
\end{equation}

In the absence of exact solution of Eq.~(\ref{eq:integral_equation}) for general $r_1$ and $r_2$, one can make progress in two limits.
One of them is a thin annulus $r_{2}-r_{1}\ll r_{2}$, which corresponds to $\ell/\sqrt{H} \ll 1$, that is to the far height-distribution tail $H\to+\infty$ for fixed $\ell$. We solve this case perturbatively in Sec.~\ref{sec:positive_tail_perturbation_theory}. The other limit, dealt with in in Sec.~\ref{section:small_hole}, is of a disk with a small hole, $r_1 \ll r_2$, which corresponds to the weakly-supercritical limit $H-H_{c}^{+}\ll H_{c}^{+}$ at $\ell \gg 1$. Instead of solving Eq.~(\ref{eq:integral_equation}) in this limit, we will directly solve Eq.~(\ref{eq:Laplace_like_t_z_r}), by a matched asymptotic expansion, and uncover a fractional-order dynamical phase transition at $H=H_c^+$.

\subsection{$\lambda H \to -\infty$}
\label{sec:positive_tail_perturbation_theory}

In the tail $H\to+\infty$ the solution of Eqs.~(\ref{eq:OFM_V_positive}) and~(\ref{eq:OFM_rho_positive}) [or of the original OFM equations~(\ref{eq:OFM_h}) and~(\ref{eq:OFM_rho})] yields $h\left(x,t\right)$ and $\rho\left(x,t\right)$ which are nearly uniform in space for most of the dynamics, except very close to $t=1$, where the dominant correction
to the uniform solution comes from the nonlinearity.
To leading order, the action is equal to that of the uniform solution~(\ref{eq:action_uniform_solution}). In this subsection we find the leading-order correction to this simple result.

A nearly spatially uniform $\rho\left(x,0\right)$ implies $\rho_{2}-\rho_{1}\ll\rho_{2}$.
Therefore, the width of the annulus, $\Delta\equiv r_{2}-r_{1}$,
is much smaller than its mean radius. Here a major simplification arises, and the height distribution can be found without solving the integral equation~(\ref{eq:integral_equation})
\footnote{We also solved the integral equation~(\ref{eq:integral_equation}) in this limit, and obtained the same results.}.
Indeed, at distances from the annulus which are much larger than the annulus' width, all of the terms in the integral in Eq.~(\ref{eq:Psi_exact_solution_as_integral}) can be taken to be constant, except for $g\left(r'\right)$. This yields the potential $\Psi$, and therefore $\tbar$, sufficiently far from the annulus:
\begin{equation}
\label{eq:t_r_z_far_regime}
\tbar\left(r,z\right)=
\frac{\Psi\left(r,z,\phi\right)}{r\cos\phi}\simeq
\frac{\bar{r}\mathcal{F}\left(r,\bar{r},z\right)}{r}\int_{r_{1}}^{r_{2}}dr'\,g\left(r'\right)
=\frac{I_{0}\bar{r}\mathcal{F}\left(r,\bar{r},z\right)}{r},\qquad\sqrt{\left(r-\bar{r}\right)^{2}+z^{2}}\gg\Delta,
\end{equation}
where
\begin{equation}
\label{eq:I0def}
I_{0}=\int_{r_{1}}^{r_{2}}dr\,g\left(r\right)
\end{equation}
is yet unknown. Surprisingly, the height distribution can be found without determining $I_0$ in terms of $\bar{r}$ and $\Delta$, and without finding $\tbar$ near the annulus, as we now show.

We begin by finding $\ell$ as a function of $I_0$.
Plugging $r=0$ into Eq.~(\ref{eq:OFM_dx_dz}) yields
\begin{equation}
\label{eq:dx_dz_on_z_axis}
\partial_{z}x\left(r=0,z\right)=2z\partial_{z}t\left(r=0,z\right).
\end{equation}
Integrating Eq.~(\ref{eq:dx_dz_on_z_axis}) with respect to $z$ from $0$ to infinity, together with the boundary conditions $x\left(r=0,z\to\infty\right)=0$ and  $x\left(r=0,z=0^{+}\right)= -\ell$, which follow from the condition~(\ref{eq:delta_initial_cond_rho}) and the assumed mirror symmetry~(\ref{eq:h_symmetric}), we obtain
 \begin{equation}
\label{eq:L_as_a_hodograph_integral}
\ell=\int_{0}^{\infty}2z\partial_{z}t\left(r=0,z\right) dz.
 \end{equation}
To obtain an approximate expression for $t=1-\bar{t}$ on the $z$ axis, we plug $r=0$ into Eq.~(\ref{eq:t_r_z_far_regime}):
 \begin{equation}
 \label{eq:t_on_z_axis}
t\left(r=0,z\right)\simeq1-\frac{\pi I_{0}\bar{r}^{2}}{\left(\bar{r}^{2}+z^{2}\right)^{3/2}}.
 \end{equation}
We now plug Eq.~(\ref{eq:t_on_z_axis}) into Eq.~(\ref{eq:L_as_a_hodograph_integral}) and obtain
\begin{equation}
\label{eq:rbar_Delta_L}
\ell \simeq 6\pi\bar{r}^{2}I_{0}\int_{0}^{\infty}\frac{z^{2}\,dz}{\left(\bar{r}^{2}+z^{2}\right)^{5/2}} = 2\pi I_{0}.
\end{equation}
A similar but a bit lengthier  calculation, which involves an integration of $t$ on the $r$ axis  (see Appendix \ref{Appendix:H_hodograph}) gives $H$ as a function of $\bar{r}$ and $I_0$:
\begin{equation}
\label{eq:H_L_rbar}
H\simeq\bar{r}^{2}\left(1+\frac{8I_{0}}{\bar{r}}\right).
\end{equation}
Next, we evaluate the action integral~(\ref{eq:action_definition}) in the $rz$ plane in terms of $\bar{r}$ and $I_0$, and then use Eqs.~(\ref{eq:rbar_Delta_L}) and~(\ref{eq:H_L_rbar}) to express the action in terms of $\ell$ and $H$ (see Appendix \ref{Appendix:action_hodograph}):
\begin{equation}
\label{eq:S_H_L_positive_tail}
S\simeq H^{2}\ell\left(1-\frac{16\ell}{3\pi\sqrt{H}}\right).
\end{equation}
The corresponding asymptotic of the scaling function $s(\alpha)$ from Eq.~(\ref{eq:action_scaling_positive_tail}) appears in the third line of Eq.~(\ref{eq:scaling_function_s_asymptotics}). Plugging Eq.~(\ref{eq:S_H_L_positive_tail}) into~(\ref{eq:action_scaling}) yields the tail
\begin{equation}
\label{eq:positive_tail_very_large_H}
-\ln\mathcal{P}\left(H,L,T\right)\simeq \frac{H^{2}L}{DT}\left(1-\frac{16L}{3\pi\sqrt{\left|\lambda\right|HT}}\right),\quad H\to+\infty
\end{equation}
in the dimensional variables.
As to be expected on physical grounds,
the probability that the interface attains height $H$ at a single point $x=0$ is larger than the probability that the height $H$ is attained uniformly on the entire ring.

We now determine the optimal history $h(x,t)$ of the height profile, conditioned on a given value of $H$, in the tail $H \to +\infty$.
In Appendix \ref{Appendix:optimal_profile_positive_tail_without_end},
we obtain the optimal height profile for the intermediate stage of the dynamics
\begin{equation}
\label{eq:optimal_profile_positive_tail_without_end}
h\left(x,t\right)\simeq H\left(1-\frac{4\ell}{\pi\sqrt{H}}\right)t+\frac{16\ell\sqrt{H}}{\pi}\exp\left[\frac{\pi\sqrt{H}}{\ell}\left(t-1\right)-2t\right]\cos\left(\frac{\pi x}{\ell}\right),\qquad\frac{\ell}{\sqrt{H}}\ll1-t\ll\frac{\ell\ln\left(\ell\sqrt{H}\right)}{\pi\sqrt{H}}.
\end{equation}
Eq.~(\ref{eq:optimal_profile_positive_tail_without_end}) is not valid at short times, before the right-hand strong inequality in Eq.~(\ref{eq:optimal_profile_positive_tail_without_end}) starts to hold, because the diffusion is not yet negligible compared with the nonlinearity.
At short times, the optimal profile is approximately uniform, $h\left(x,t\right)\simeq H\left[1-4\ell/\left(\pi\sqrt{H}\right)\right]t$, with a very small diffusion-dominated correction, whose contribution to the action is negligible in the leading order.
Eq.~(\ref{eq:optimal_profile_positive_tail_without_end}) also breaks down in a narrow boundary layer $1-t\sim \ell/\sqrt{H}$, where higher harmonics become important. Although finding $h(x,t)$ explicitly in this boundary layer is difficult, it is relatively straightforward to obtain the optimal interface profile at $t=1$ (see Appendix \ref{Appendix:final_optimal_profile_positive_tail}):
\begin{equation}
\label{eq:final_optimal_profile_positive_tail}
h\left(x,t=1\right)\simeq H\left[1-\frac{2\ell}{\sqrt{H}}\sqrt{\frac{\left|x\right|}{\ell}\left(2-\frac{\left|x\right|}{\ell}\right)}\,\right].
\end{equation}
This profile contains a cusp singularity at $x=0$, as does its counterpart in an infinite system \citep{MKV}. Actually, this singularity is smoothened out by the diffusion (which we neglected in this and the preceding section) over a thin (much narrower than $\ell$) boundary layer around $x=0$, whose contribution to the action is negligible.

Evaluating the nonlinear term $\left(\partial_{x}h\right)^{2}$ and the diffusion term $\partial_{x}^{2}h$ from Eq.~(\ref{eq:KPZ_dimensionless}) on the profile~(\ref{eq:final_optimal_profile_positive_tail}), we find that the former term dominates
for $H\gg \ell^{-2}$ (except in the narrow boundary layer around $x=0$).
Alongside with the thin-annulus condition $H\gg \ell^2$, the applicability condition of the results of this subsection can be written as
\begin{equation}
\label{eq:positive_h_tail_limit}
H\gg\max\left\{ \ell^{2},\ell^{-2}\right\}.
\end{equation}

\subsection{A fractional phase transition at large $L/\sqrt{t}$}
\label{section:small_hole}

In this subsection we consider the weakly-supercritical inviscid limit $0<H-H_{c}^{+}\ll H_{c}^{+}$ at $\ell \gg 1$. In this limit the infinite-system solution almost, but not quite, fits into the ring. A weak supercriticality implies $r_1 \ll r_2$.
In the electrostatic analogy of Sec.~\ref{sec:electrostatic}
this corresponds to finding the potential where the Dirichlet boundary condition~(\ref{eq:Psi_Dirichlet_condition}) is specified on a disk with a small hole. Here one can directly solve the problem~(\ref{eq:Laplace_like_t_z_r})-(\ref{eq:t_boundary_condition_at_infinity}) using matched asymptotic expansions, see Appendix \ref{appendix:small_hole_matched_asymptotics}. Then a calculation similar to that described in the previous subsection (see Appendix \ref{appendix:small_hole_ell_H_S}) gives
\begin{eqnarray}
\label{eq:ell_r1_r2_main_text}
\ell&\simeq&\frac{4r_{2}}{\pi}\left(1-\frac{r_{1}^{2}}{3r_{2}^{2}}\right),\\
\label{eq:H_r2_main_text}
H&\simeq&2r_{2}^{2}.
\end{eqnarray}
From Eqs.~(\ref{eq:Hcplus_def}),~(\ref{eq:ell_r1_r2_main_text}) and~(\ref{eq:H_r2_main_text}) we obtain
\begin{equation}
\label{eq:H_Hcplus_r_1}
H-H_{c}^{+}\simeq\frac{4r_{1}^{2}}{3}.
\end{equation}
The action can be calculated in terms of $r_1$ and $r_2$.
In the zeroth order, $r_1 = 0$, we obtain
\begin{equation}
\label{eq:S_Y2_zero_order}
S\left(r_{1}=0,r_{2}\right)=\frac{64r_{2}^{5}}{15\pi}=\frac{8\sqrt{2}\,H^{5/2}}{15\pi},
\end{equation}
reproducing the infinite-system result~(\ref{eq:positive_tail_no_finite_size_effects}).
In order to evaluate the small correction for $r_1>0$, it is necessary to find the correction to our solution to the problem~(\ref{eq:Laplace_like_t_z_r})-(\ref{eq:t_boundary_condition_at_infinity}).
As we argue in Appendix \ref{appendix:small_hole_ell_H_S}, the scaling behavior of the correction to the action is
\begin{equation}
\label{eq:S_Y1_scaling}
\Delta S\left(r_{1},r_{2}\right)=S\left(r_{1},r_{2}\right)-S\left(r_{1}=0,r_{2}\right)\simeq\tilde{c}_{1}r_{1}^{5},
\end{equation}
where $\tilde{c}_{1}$ is a constant of order unity. Combining Eq.~(\ref{eq:S_Y1_scaling}) with Eqs.~(\ref{eq:ell_r1_r2_main_text}) and~(\ref{eq:H_r2_main_text})
we obtain the third line of Eq.~(\ref{eq:scaling_function_s_asymptotics}),
with $c_{2}=-\tilde{c}_{1}\left[3\pi/\left(4\sqrt{2}\right)\right]^{5/2}$.
Correspondingly, the action is given, close to criticality, by
\begin{equation}
\label{eq:S_H_Hc_scaling}
S\left(H,\ell\right)\simeq\begin{cases}
S_{2}\left(H,\ell\right)+\frac{\left(H-H_{c}^{+}\right)^{3}}{3\pi\sqrt{2H_{c}^{+}}}, & 1\ll H<H_{c}^{+},\quad H_{c}^{+}-H\ll H_{c}^{+},\\
S_{2}\left(H,\ell\right)-c_{1}\left(H-H_{c}^{+}\right)^{5/2}, & 1\ll H_{c}^{+}<H,\quad H-H_{c}^{+}\ll H_{c}^{+},
\end{cases}
\end{equation}
where $c_{1}=-\left(3/4\right)^{5/2}\tilde{c_1}$,
\begin{equation}
S_{2}\left(H,\ell\right)=\frac{\sqrt{2H_{c}^{+}}}{\pi}\left[\frac{8\left(H_{c}^{+}\right)^{2}}{15}+\frac{4H_{c}^{+}\left(H-H_{c}^{+}\right)}{3}+\left(H-H_{c}^{+}\right)^{2}\right],
\end{equation}
and the $\ell$ dependence is through Eq.~(\ref{eq:Hcplus_def}).
Eq.~(\ref{eq:S_H_Hc_scaling}) implies that $S$ experiences a fractional-order phase transition \citep{Hilfer}, of order $5/2$, at $H=H_c^+$.
Instead of obtaining $c_1$ via a cumbersome analytic calculation (which is briefly outlined in Appendix \ref{appendix:small_hole_ell_H_S}),
we solved the problem~(\ref{eq:Laplace_like_t_z_r})-(\ref{eq:t_boundary_condition_at_infinity}) numerically, using an artificial relaxation method, and evaluated the action numerically. The resulting data, shown in Fig.~\ref{fig:smallHole}, confirm Eq.~(\ref{eq:S_Y1_scaling}) with $\tilde{c}_1 \simeq - 0.14$. This leads to Eq.~(\ref{eq:S_H_Hc_scaling}) with $c_1 \simeq0.068$ and to $c_2 \simeq 0.50$ in the third line of Eq.~(\ref{eq:scaling_function_s_asymptotics}). By virtue of the negative sign of the $c_1$ term in Eq.~(\ref{eq:S_H_Hc_scaling}), finite-$L$ effects increase the probability of observing a height $H>H_c^+$. The transition at $H=H_c^+$ is of order $5/2$, so the finite-$L$ effects are relatively weak for weakly-supercritical $H$. Note that this phase transition (a non-analytic behavior of $S$  at $H=H_c^+$) is obtained only
in the inviscid limit. It is smoothed out by the diffusion in the region $|H-H_c^+|\sim \ell$. As $\ell$ increases, the
relative width of the transition region, $|H-H_c^+|/H_c^+ \sim 1/\ell$, goes down, and the transition becomes increasingly sharper.

\begin{figure}[ht]
\includegraphics[width=0.5\textwidth,clip=]{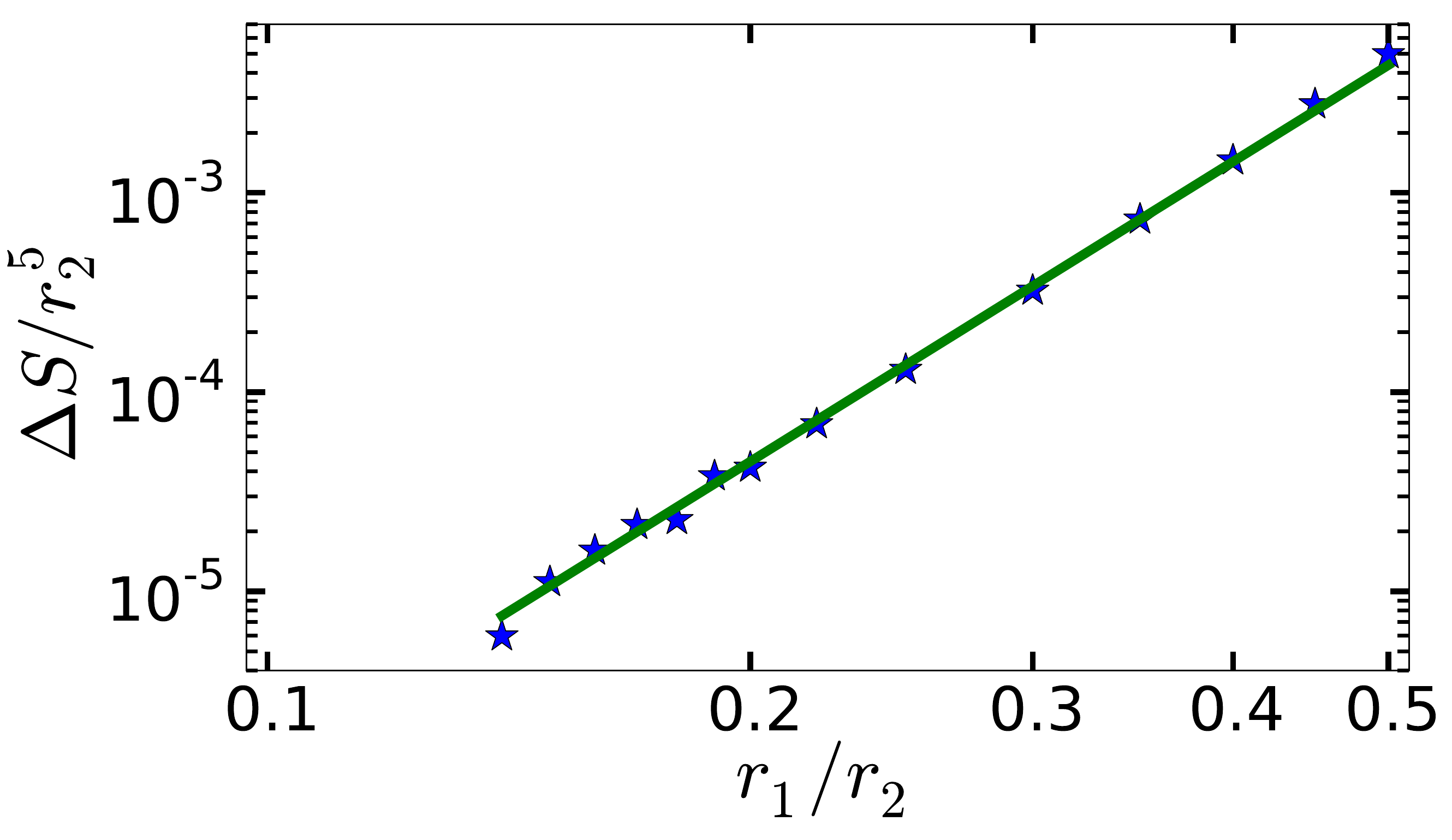}
\caption{$\Delta S /r_{2}^{5}$ as a function of $r_1 / r_2$, see Eq.~(\ref{eq:S_Y1_scaling}). Shown on a log-log scale are data from numerical solutions of Eq.~(\ref{eq:Laplace_like_t_z_r}) (markers) and the fit to $|\tilde{c}_1|\left(r_{1}/r_{2}\right)^{5}$ (solid line), where $\tilde{c}_1\simeq - 0.14$.}
\label{fig:smallHole}
\end{figure}

\section{Summary and discussion}
\label{disc}

We have employed the optimal fluctuation method (OFM) to study the interface-height distribution $\mathcal{P}(H,L,t)$ for the KPZ equation in a ring $|x| < L$  at short times $t\ll\nu^{5}/D^{2}\lambda^{4}$, when the stochastic process starts from a flat interface.
We have shown that, in a proper moving frame\footnote{See footnote \ref{footnote:displacement}.}
and at fixed $L/\sqrt{t}$, the large deviation function of the height $S=- \epsilon \,\ln \mathcal{P}(H,L,t)$ is given, as a function of $H$, by Eq.~(\ref{eq:Stilde_L}) in the body of the distribution; the tails are described by Eqs.~(\ref{eq:MKV_TW_action})
and~(\ref{eq:S_H_L_positive_tail}).
The height distribution is plotted schematically in Figure~\ref{fig:S_schematic}, and
a phase diagram of the system in the $(L/\sqrt{t}, H)$ plane is shown in Figure \ref{fig:phase_diagram}.
Because of the smallness of the parameter $\epsilon=D\lambda^{2}\sqrt{t}/\nu^{5/2}$ the OFM results are asymptotically exact.

For $L\gg\sqrt{t}$, there are no finite-size effects on the  $\lambda H \to \infty$ tail of the height distribution. This far tail, of the form $-\ln\mathcal{P} \sim \left|H\right|^{3/2} \! / \, t^{1/2}$, coincides with the slower-decaying tail of the GOE Tracy-Widom distribution \citep{TracyWidom1996} as it does at $L=\infty$. This tail was previously found in a whole class of infinite one-dimensional settings with  deterministic initial conditions: both for long times \citep{CLD, Sasamoto2010,Calabrese2010,Dotsenko2010,Amir2011} and for short times \cite{KK2007,MKV,DMRS,KMSparabola}.
As we have shown here, it persists in finite systems as well.
The optimal history of the interface, conditioned on reaching a large height $\lambda H > 0$ is a traveling $h$-front driven by a stationary profile of $\rho$, the optimal realization of the noise.
For $L\ll\sqrt{t}$ we have found a sharp transition around a critical value $H=H_{c}^{-}\left(L / \sqrt{t} \right)$. In the limit $L/\sqrt{t} \to 0$, the transition has the character of a mean-field-like second-order phase transition, as it corresponds to a singularity of the large deviation function $S$. For subcritical $H$ the optimal profile of $h$ is (almost) spatially uniform. For supercritical $H$ the optimal profile of $h$ is non-uniform, and $S$ is given in a parametric form by Eqs.~(\ref{eq:c_as_function_of_kappa}) and~(\ref{eq:S_as_function_of_kappa_and_L}), again yielding the behavior $-\ln\mathcal{P} \sim \left|H\right|^{3/2} \! / \, t^{1/2}$ at the tail $\lambda H \to +\infty$.

The $\lambda H \to -\infty$ tail of the height distribution is strongly affected by the finiteness of the system: to leading order it is given by $-\ln\mathcal{P} \sim H^2 L / t$, in contrast to $-\ln\mathcal{P} \sim H^{5/2} \! / \, t^{1/2}$ for an infinite system \citep{KK2007,MKV}.
This leading-order contribution to the height distribution is due to the ``cost" of lifting the entire interface uniformly to height $H$. We found the first-order correction, of order $H^{3/2}L^{2} / t$, to this simple result. The corresponding optimal trajectory is uniform for most of the dynamics, but towards the end it becomes strongly nonuniform, until a cusp singularity is formed at the observation time.
For $L \gg \sqrt{t}$, the $\lambda H \to -\infty$ tail of the height distribution has a double structure: its behavior changes from Eq.~(\ref{eq:positive_tail_no_finite_size_effects}) at intermediately large $H$ (coinciding with that of an infinite system) to Eq.~(\ref{eq:positive_tail_very_large_H}) at very large $H$. The transition between these two regimes is sharp. If the diffusion is neglected (that is if $L/\sqrt{t}$ is sufficiently large), it has the character of a fractional-order phase transition of order $5/2$, at a critical value $H=H_{c}^{+}\left(L / \sqrt{t} \right)$.

As we have shown here, a strong asymmetry between the two tails of $\mathcal{P}(H)$, observed in an infinite system, persists in the finite system. To emphasize the asymmetry, it is useful to reinterpret some of our results by following the evolution  of $\mathcal{P}(H)$ in time at fixed $L$. As is evident from Fig.~\ref{fig:S_schematic} (a), at short times the height distribution is the same as in an infinite system except at very large positive $H$, where the finite-size effects
cause the fractional phase-transition  of order $5/2$, at a critical value $H=H_{c}^{+}\left(L / \sqrt{t} \right)$, and create the Gaussian far right tail $\sim H^2 L/t$ at still larger positive heights. As time increases, this Gaussian tail propagates from right to left. First, it eliminates the phase transition at $H=H_{c}^{+}\left(L / \sqrt{t} \right)$. Then it wipes out the $H^{5/2}/t^{1/2}$ tail. Later on it changes the character of the Gaussian central part of the distribution and causes the sharp transition
around the critical value $H=H_{c}^{-}\left(L / \sqrt{t} \right)$. At still larger times, it continues ``pushing" to the left, establishing its dominance over a progressively larger part of the distribution\footnote{Here we assume that time is not too long, so as the WNT is still applicable. This can always be achieved when the noise magnitude $D$ is sufficiently small.}.

It would be interesting to determine whether any of these short-time results  persist at longer times.
The OFM can be also used for studying short-time statistics of other quantities pertaining to the KPZ equation. One interesting quantity is the spatial-average interface height, $(2L)^{-1} \int_{-L}^{L} h(x,t)\,dx$. For this quantity, as one can see,  the sharp transition at $H=H_c^-$, that we observed in this paper, gives way to a true second-order dynamic phase transition even at finite values of $L/\sqrt{T}$.

\section*{Acknowledgements}

We thank Tal Agranov for useful discussions. We
acknowledge support from the Israel Science Foundation (Grant No. 807/16). N.R.S. also acknowledges support from the Hoffman Leadership and Responsibility Fellowship.

\begin{appendices}


\section{Derivation of the OFM equations}
\label{Appendix:OFMderivation}

\renewcommand{\theequation}{A\arabic{equation}}
\setcounter{equation}{0}

For completeness, here we present a brief derivation of the OFM equations and boundary conditions. Isolating the noise term from Eq.~(\ref{eq:KPZ_dimensional}), we obtain
\begin{equation}\label{actn0}
\sqrt{D}\,\xi(x,t)=\partial_{t} h-\nu \partial_{x}^2 h-\frac{\lambda}{2} \left(\partial_{x} h\right)^2.
\end{equation}
The probability of a history $h(x,t)$ of the interface height is therefore $-\ln\mathcal{P}\left[h\left(x,t\right)\right]\simeq S/D$, where
\begin{equation}\label{actn}
S=\frac{1}{2}\int_{0}^{T}dt\int_{-L}^{L}dx \left[\partial_{t} h-\nu \partial_{x}^2 h-\frac{\lambda}{2} \left(\partial_{x} h\right)^2\right]^2.
\end{equation}
In the weak-noise limit the main contribution to the probability $\mathcal{P}\left(H,L,T\right)$ of observing an unusually large value of $H$ comes from the ``optimal path" $h(x,t)$ that minimizes $S$. The variation of $S$ is
\begin{equation}
\label{variation}
\delta S=\int_{0}^{T}dt\int_{-L}^{L}dx\left[\partial_{t}h-\nu\partial_{x}^{2}h-\frac{\lambda}{2}\left(\partial_{x}h\right)^{2}\right]\left(\partial_{t}
\delta h-\nu\partial_{x}^{2}\delta h-\lambda\partial_{x}h\,\partial_{x}\delta h\right).
\end{equation}
By analogy with classical mechanics, one can introduce the ``momentum density" field $\rho(x,t)=\delta \mathcal{L}/\delta v$, where $v\equiv \partial_t h$, and
\begin{equation}
\mathcal{L}\{h\}=\frac{1}{2}\int_{-L}^{L}dx \left[\partial_{t} h-\nu \partial_{x}^2 h-\frac{\lambda}{2} \left(\partial_{x} h\right)^2\right]^2
\end{equation}
is the Lagrangian.  In this way we obtain
\begin{equation}\label{heqA}
\partial_{t}h=\nu \partial_{x}^2 h +\frac{\lambda}{2} \left(\partial_x h\right)^2+\rho,
\end{equation}
the first of the two Hamilton equations. Rewriting the variation~(\ref{variation}) as
\begin{equation}
\delta S=\int_{0}^{T}dt\int_{-L}^{L}dx\,\rho \,(\partial_{t}\delta h-\nu \partial_{x}^2\delta h -\lambda \partial_x h \,\partial_x \delta h),
\end{equation}
and integrating by parts, we arrive at the second Hamilton equation:
\begin{equation}\label{peqA}
\partial_{t}\rho=-\nu \partial_{x}^2 \rho +\lambda \partial_x \left(\rho \partial_x h\right).
\end{equation}
The boundary terms in $x$, emerging in the integrations by parts, vanish because of the periodic boundary conditions on $x$. There also appear two boundary terms in time: at $t=0$ and $t=T$.
The boundary term  $\int dx  \,\rho(x,0) \,\delta h(x,0)$ vanishes because the height profile at $t=0$ is fixed by the flat initial condition~(\ref{eq:flat_initial_cond}).
In order to determine the boundary condition at $t=T$, we add a Lagrange multiplier,
\begin{equation}
\label{eq:LagrangeMultiplierDef}
\Lambda h\left(0,T\right)=\Lambda\int_{-L}^{L}\delta\left(x\right)h\left(x,T\right)dx,
\end{equation}
to the action, due to the constraint $h(0,T) = H$. Taking the term~(\ref{eq:LagrangeMultiplierDef}) into account, the boundary term of $\delta S$ at $t=T$ becomes $\int dx\,\left[\rho(x,T)-\Lambda\delta\left(x\right)\right]\,\delta h(x,T)$. This leads to the boundary condition \cite{KK2007,MKV}
\begin{equation}\label{pTA}
    \rho(x,T)=\Lambda \,\delta(x).
\end{equation}
$\Lambda$ is found by the condition $h(x=0,T)=H$. Upon the rescaling $t/T\to t$, $x/\sqrt{\nu T} \to x$, $|\lambda|h/\nu\to h$ and  $|\lambda|T\rho/\nu\to \rho$, one arrives
at Eqs.~(\ref{eq:OFM_h})--(\ref{eq:action_definition}) of the main text, with rescaled $H$
and $\Lambda$.

\section{Traveling $h$-front solution}
\label{Appendix:TW}
\renewcommand{\theequation}{B\arabic{equation}}
\setcounter{equation}{0}

The Hamiltonian~(\ref{eq:hamiltonian_hh_def}) is a constant of motion, so
\begin{equation}
\label{eq:hamiltonian_hh}
\mathfrak{h}=\frac{\rho_{0}\left(V_{0}^{2}-\rho_{0}-2c\right)}{2}=\EE=\text{const},
\end{equation}
where $\EE$ is the ``energy''.
Using Eq.~(\ref{eq:hamiltonian_hh}), we express $V_0$ in terms of $\rho_0$, $c$ and $\EE$. Plugging this expression for $V_0$ into Eq.~(\ref{eq:OFM_TW_rho0}) (with $C_1=0$) gives a single first-order equation
\begin{equation}
\label{eq:rho0_c_EE}
\rho_{0}'=\pm\sqrt{\rho_{0}^{3}+2c\rho_{0}^{2}+2\EE\rho_{0}}
\end{equation}
which appears in countless applications. One can rewrite Eq.~(\ref{eq:rho0_c_EE}) as an equation of motion for a classical particle of unit mass and zero energy,
\begin{equation}
\label{eq:rho0_mechanical_analogy}
\frac{\left(\rho_{0}'\right)^{2}}{2}+U\left(\rho_{0}\right)=0,
\end{equation}
where
\begin{equation}
\label{eq:U0def}
U\left(\rho_{0}\right)=-\left(\frac{\rho_{0}^{3}}{2}+c\rho_{0}^{2}+\EE\rho_{0}\right)
\end{equation}
is the effective potential. Reasonable solutions are obtained when $c>0$  and $0<2\EE<c^{2}$. The potential $U\left(\rho_{0}\right)$ is depicted in Fig. \ref{fig:U0}.

\begin{figure}[ht]
\includegraphics[width=0.4\textwidth,clip=]{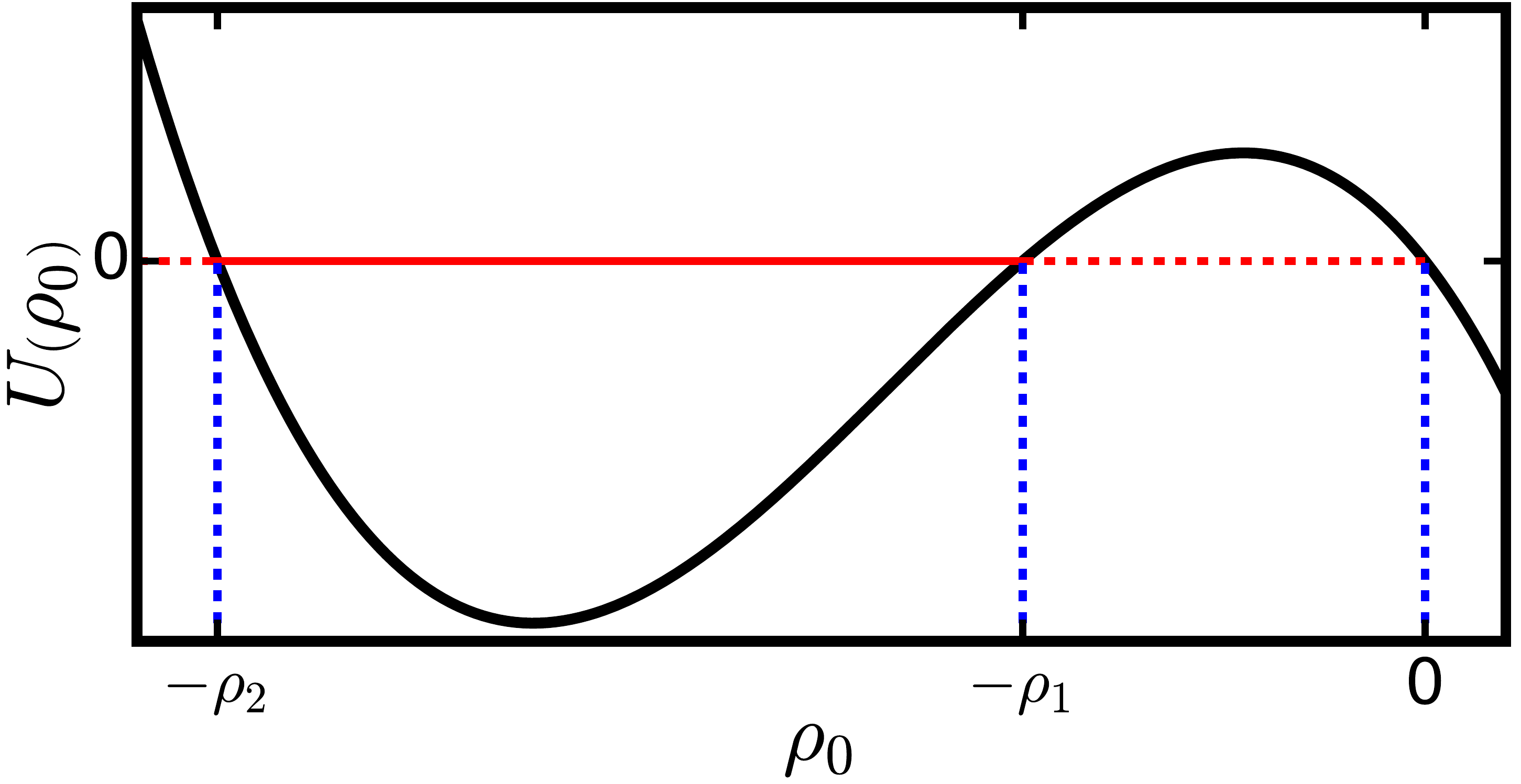}
\caption{The effective potential~(\ref{eq:U0def}). $U=0$ at the points $-\rho_2$, $-\rho_1$ and 0. The oscillations occur in the region $-\rho_2 < \rho_0 < -\rho_1$. In the weakly supercritical regime, $\rho_1$ and $\rho_2$ approach the minimum of the potential, and the motion becomes nearly harmonic. In the strongly supercritical
regime~(\ref{eq:strongly_nonlinear_limit}) $\rho_1$ approaches 0, and one observes a soliton-like solution for $\rho_0(x)$.}
\label{fig:U0}
\end{figure}

From the assumed mirror symmetry~(\ref{eq:h_symmetric}), $\rho_{0}'$ must vanish at $x=0$ and at $x=\pm \ell $. It then follows, from Eq.~(\ref{eq:rho0_mechanical_analogy}), that $U\left[\rho_{0}\left(0\right)\right]=U\left[\rho_{0}\left(\pm \ell\right)\right]=0$. Therefore the values of $\rho_0$ at $x=0$ and at $x=\pm \ell$, which we denote $-\rho_2$ and $-\rho_1$, respectively, are immediately found from Eq.~(\ref{eq:U0def}):
\begin{eqnarray}
\label{eq:rho2def}
\rho_{0}\left(0\right)&=&-c-\sqrt{c^{2}-2\EE}\equiv-\rho_{2},\\
\label{eq:rho1def}
\rho_{0}\left(\pm \ell \right)&=&-c+\sqrt{c^{2}-2\EE}\equiv-\rho_{1}.
\end{eqnarray}
Rewriting Eq.~(\ref{eq:rho0_c_EE}) in terms of $\rho_1$ and $\rho_2$, we obtain
\begin{equation}
\label{eq:rho0prime_rho1_rho2}
\rho_{0}'=\pm\sqrt{\rho_{0}\left(\rho_{0}+\rho_{1}\right)\left(\rho_{0}+\rho_{2}\right)}.
\end{equation}
Eq.~(\ref{eq:rho0prime_rho1_rho2}) can be integrated to yield the solution in terms of the Jacobi elliptic function $\text{dn}(u, m )$ \cite{elliptic_functions}:
\begin{equation}
\label{eq:rho0sol_with_rho2}
\rho_{0}\left(x\right)=-\rho_{2} \, \text{dn}^{2}\left(\frac{\sqrt{\rho_{2}}\,x}{2},\, m \right),
\end{equation}
where
\begin{equation}
\label{eq:mdef}
 m =1-\frac{\rho_{1}}{\rho_{2}}
\end{equation}
is the elliptic modulus. We have omitted in Eq.~(\ref{eq:rho0sol_with_rho2}) an arbitrary constant, resulting from the translational symmetry of the solution, and thus have set the minimum of $\rho_0$ to be at $x=0$.

The periodic boundary conditions~(\ref{boundary_cond}) give rise to the constraint that the system length must contain an integer number of periods of the oscillating function $\rho$. However, the action of the `fundamental mode'  is minimal (see Appendix \ref{multiple_waves}),
so we demand
\begin{equation}
\label{eq:rho2_as_function_of_kappa}
\rho_{2}=\left[\frac{2 \EllipticK \left( m \right)}{\ell}\right]^{2},
 \end{equation}
where $\EllipticK$  is the complete elliptic integral of the first kind \citep{elliptic_functions}.
We use Eq.~(\ref{eq:rho2_as_function_of_kappa}) in order to re-parametrize our results in terms of $ m $ and $\ell$.
Plugging Eq.~(\ref{eq:rho2_as_function_of_kappa}) into Eq.~(\ref{eq:rho0sol_with_rho2}) yields Eq.~(\ref{eq:rho0_negative_tail_exact}), and plugging Eq.~(\ref{eq:rho2_as_function_of_kappa}) into~(\ref{eq:mdef}) yields
\begin{equation}
\label{eq:rho1_as_function_of_kappa}
\rho_{1} = \left(1- m \right)\rho_{2}=\left(1- m \right)\left[\frac{2 \EllipticK \left( m \right)}{ \ell }\right]^{2}.
\end{equation}
The condition $h\left(0,1\right)=H$ implies that $c\simeq\left|H\right|$. On the other hand, $c$ can be expressed via $ m $ and $\ell$, by plugging Eqs.~(\ref{eq:rho2_as_function_of_kappa}) and~(\ref{eq:rho1_as_function_of_kappa}) into the relation $c=\left(\rho_{1}+\rho_{2}\right)/2$, which follows from Eqs.~(\ref{eq:rho2def}) and~(\ref{eq:rho1def}).
This yields Eq.~(\ref{eq:c_as_function_of_kappa}).
We now find the optimal interface profile. Plugging Eq.~(\ref{eq:rho0prime_rho1_rho2}) into Eq.~(\ref{eq:OFM_TW_rho0}), we obtain:
\begin{equation}
\label{eq:V0_as_function_of_rho0_1_2}
V_{0}(x)=\text{sgn}\left(x\right)\sqrt{\frac{\left[\rho_{0}(x)+\rho_{1}\right]\left[\rho_{0}(x)+\rho_{2}\right]}{\rho_{0}(x)}}.
\end{equation}
We chose the sign in Eq.~(\ref{eq:V0_as_function_of_rho0_1_2}) such that
the minimum of $h_0(x)$ is attained at $x=0$.
In order to obtain $V_0(x)$ in terms of $ m $ and $\ell$, we plug Eqs.~(\ref{eq:rho2_as_function_of_kappa}),~(\ref{eq:rho1_as_function_of_kappa}) and~(\ref{eq:rho0_negative_tail_exact}) into Eq.~(\ref{eq:V0_as_function_of_rho0_1_2}), which, with the use of elliptic function identities, yields  Eq.~(\ref{eq:V0_as_function_of_rho0_and_kappa}).

The traveling $h$-front solution found here is only valid for $|H|$ larger than a critical value, as discussed in the main text.

\section{Higher harmonics are not allowed}
\label{multiple_waves}
\renewcommand{\theequation}{C\arabic{equation}}
\setcounter{equation}{0}

Let us return to the derivation in Appendix \ref{Appendix:TW}, but assume that the solution represents a mode with $n>1$ wavelengths, where $n$ is an integer. Now Eq.~(\ref{eq:rho2_as_function_of_kappa}) becomes
\begin{equation}
\label{eq:rho2_as_function_of_kappa_n}
\rho_{2}=\left[\frac{2n \EllipticK \left( m \right)}{\ell}\right]^{2}.
\end{equation}
This yields
\begin{eqnarray}
\rho_{1}&=&\left(1- m \right)\rho_{2}=\left(1- m \right)\left[\frac{2 n \EllipticK \left( m \right)}{ \ell }\right]^{2}, \\
\left|H\right|\simeq c&=&\frac{\rho_{1}+\rho_{2}}{2} =2\left(2- m \right)\left[\frac{ n \EllipticK \left( m \right)}{ \ell }\right]^{2}, \\
\rho_{0}\left(x\right) &=& -\left[\frac{2 n \EllipticK\left( m \right)}{\ell}\right]^{2}\text{dn}^{2}\left[\frac{n \EllipticK\left( m \right)x}{\ell},\, m \right].
\end{eqnarray}
Generalizing Eq.~(\ref{eq:Hcminus_def}), the critical value of $H$ for the existence of the $n$-th
mode is equal to $n^2 H_c^-$. Therefore, at $\left|H_{c}^{-}\right|<\left|H\right|<4\left|H_{c}^{-}\right|$  only the fundamental mode $n=1$ is possible. At $4\left|H_{c}^{-}\right|<\left|H\right|<9\left|H_{c}^{-}\right|$ the only possible modes are $n=1$ and $2$. At $9\left|H_{c}^{-}\right|<\left|H\right|<16\left|H_{c}^{-}\right|$ there are three possible modes: $n=1$, $2$ and $3$, etc. We now compare the action for the $n$-th mode, which we denote $S_n$, with $S_1$:
\begin{equation}
S_{n}\left(H,\ell\right)=\frac{1}{2}\int_{-\ell}^{\ell}dx\,\rho_{0}^{2}\left(x\right)=nS_{1}\left(H,\ell/n\right).
\end{equation}
For $\left|H\right|\gg\left|H_{c}^{-}\right|$, we have $S_{1}\left(H,\ell\right)\simeq\frac{8\sqrt{2}}{3}\left|H\right|^{3/2}$, see Eq.~(\ref{eq:MKV_TW_action}), so $S_{n}\left(H,\ell\right) = nS_{1}\left(H,\ell\right)$ for $n>1$ in this limit.
  As we checked numerically, the inequality $S_{n}\left(H,\ell\right) > S_{1}\left(H,\ell\right)$ holds at all supercritical values of $H$, as shown in Fig.~\ref{fig:shorterTW} for $n=2$ and $3$. Therefore, the higher harmonics are not allowed.
A similar situation occurs in other large-deviation problems with periodic boundaries, e.g. \cite{Zarfaty2016}.

\begin{figure}[ht]
\includegraphics[width=0.5\textwidth,clip=]{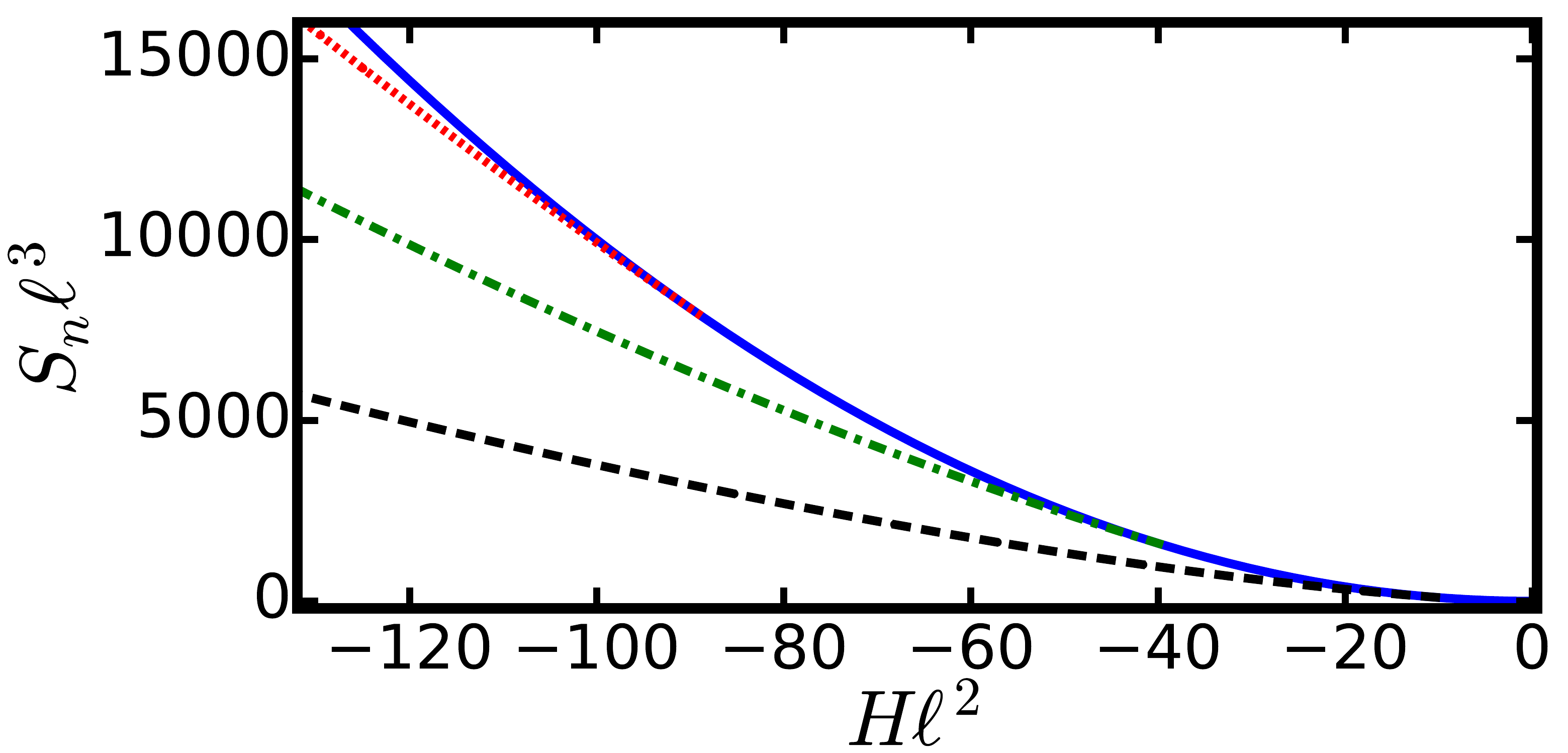}
\caption{The action $S_{n} \ell^3$ of the $n$th traveling $h$-front mode versus $H\ell^{2}$ for the three first modes: the fundamental mode $n=1$  (dashed), the second harmonic $n=2$ (dot-dashed) and the third harmonic $n=3$ (dotted). The action of the uniform solution is also shown, in solid. As one can see, for $\left|H\right|>\left|H_{c}^{-}\right|$  the action is minimal for $n=1$.}
\label{fig:shorterTW}
\end{figure}

\section{Evaluating the action for $\lambda H \to -\infty$}

\renewcommand{\theequation}{D\arabic{equation}}
\setcounter{equation}{0}

In this Appendix we evaluate the (rescaled) $H$ and $S$ in the thin-annulus limit $\Delta \ll \bar{r}$ as functions of the mean radius of the annulus, $\bar{r}$, and of $I_0$~(\ref{eq:I0def}). We then use these results, together with Eq.~(\ref{eq:rbar_Delta_L}), to evaluate the action $S$ as a function of $H$ and $\ell$.

\subsubsection{Finding $H$ as a function of $\bar{r}$ and $I_0$}
\label{Appendix:H_hodograph}
It follows from the symmetry~(\ref{eq:h_symmetric}) that $2z\left(x=0,t\right)=V\left(x=0,t\right)=\partial_{x}h\left(x=0,t\right)=0$ at all times $0<t<1$. Therefore, it can be seen from Eq.~(\ref{eq:OFM_h}) (after neglecting the diffusion term) that the interface height at $t=1$ is given by
\begin{equation}
H=h\left(x=0,t=1\right)=\int_{0}^{1}\rho\left(x=0,t\right)dt.
\end{equation}
We can calculate this integral in the $rz$ plane. Integrating by parts, we obtain:
\begin{equation}
\label{eq:H_integral_rz_plane}
H=\int_{0}^{1}r^{2}\left(x=0,t\right)dt=-\int_{r_{2}}^{\infty}r^{2}\frac{\partial\tbar\left(z=0,r\right)}{\partial r}dr=-\left[\tbar\left(z=0,r\right)r^{2}\right]_{r=r_{2}}^{\infty}+\int_{r_{2}}^{\infty}2r\tbar\left(z=0,r\right)dr.
\end{equation}
The boundary term at $r\to\infty$ in Eq.~(\ref{eq:H_integral_rz_plane}) vanishes due to the asymptotic behavior~(\ref{eq:tbar_asymptotics_large_rz}) of $\tbar$. It is immediately seen from the Dirichlet boundary condition~(\ref{eq:t_Dirichlet_condition}) that the boundary term at $r=r_2$ is equal to $r_2^2 = \rho_2 \simeq \bar{r}^2$. This term corresponds to the uniform solution~(\ref{eq:uniform_solution}).
We now evaluate the final integral in Eq.~(\ref{eq:H_integral_rz_plane}), which gives the leading-order correction to the uniform result.
The leading contribution to this integral is from the region $r-\bar{r}\gg\Delta$, where Eq.~(\ref{eq:t_r_z_far_regime}) is valid
\footnote{The contribution to the integral in Eq.~(\ref{eq:H_integral_rz_plane}) from the region $r-\bar{r} =O(\Delta)$ can be estimated by replacing $r$ by $\bar{r}$ and $\tbar$ by $1$ in the integrand, and introducing an upper bound of order $r_{2}+O\left(\Delta\right)$. The result is of order $O\left(\bar{r}\Delta\right)$, which, in the thin annulus limit $\Delta \ll \bar{r}$, is much smaller than the contribution~(\ref{eq:H_far_evaluation}) coming from the region $r-\bar{r}\gg\Delta$.}.
We plug Eq.~(\ref{eq:t_r_z_far_regime}) into the final integral in Eq.~(\ref{eq:H_integral_rz_plane}), replace the lower bound of the integral by $\bar{r}$ and obtain:
\begin{eqnarray}
\label{eq:H_far_evaluation}
\!\!\!\!\!\!\!\!\!\!\!\!\int_{r_{2}}^{\infty}2r\tbar\left(z=0,r\right)dr&\simeq&2I_{0}\bar{r}\int_{\bar{r}}^{\infty}\mathcal{F}\left(r,\bar{r},0\right)dr\nonumber\\
&=&4I_{0}\bar{r}\int_{1}^{\infty}d\tilde{r}\frac{\left(\tilde{r}^{2}+1\right)\EllipticK\left[-\frac{4\tilde{r}}{\left(\tilde{r}-1\right)^{2}}\right]-\left(\tilde{r}-1\right)^{2}\EllipticE\left[-\frac{4\tilde{r}}{\left(\tilde{r}-1\right)^{2}}\right]}{\tilde{r}\left(\tilde{r}-1\right)}\nonumber\\
&=&4I_{0}\bar{r}\int_{0}^{\infty}dw\frac{\left(w+2+2\sqrt{w+1}\right)\left[\left(w+2\right)\EllipticK\left(-w\right)-2\EllipticE\left(-w\right)\right]}{w^{2}\left(w+1+\sqrt{w+1}\right)}=8I_{0}\bar{r}.
\end{eqnarray}
This equation, together with the boundary term which we already calculated, yields Eq.~(\ref{eq:H_L_rbar}).

\subsubsection{Evaluating the action}
\label{Appendix:action_hodograph}

We first transform the action integral~(\ref{eq:action_definition}) to the $rz$ plane:
\begin{equation}
\label{eq:action_rz_plane_with_Jacobian}
S=\frac{1}{2}\int_{0}^{1}dt\int_{-\ell}^{\ell}dx\,\rho^{2}\left(x,t\right)=\frac{1}{2}\int_{0}^{1}dt\int_{-\ell}^{\ell}dx\,r^{4}\left(x,t\right)=\frac{1}{2}\int_{-\infty}^{\infty}dz\int_{0}^{\infty}dr\left|J\left(r,z\right)\right|r^{4},
\end{equation}
where
\begin{equation}
\label{eq:Jacobian_def}
J\left(r,z\right)=\left|\begin{array}{cc}
\partial_{z}t & \partial_{r}t\\
\partial_{z}x & \partial_{r}x
\end{array}\right|=\partial_{r}x\partial_{z}t-\partial_{z}x\partial_{r}t
\end{equation}
is the Jacobian of the transformation. We cannot evaluate $J$ directly from Eq.~(\ref{eq:Jacobian_def}), because we do not have an explicit expression for $x(r,z)$. Instead, we plug Eqs.~(\ref{eq:OFM_dx_dz}) and~(\ref{eq:OFM_dx_dr}) into Eq.~(\ref{eq:Jacobian_def}) in order to eliminate $x$ from the equation, and obtain an expression for the Jacobian which only involves derivatives of $t(r,z)$:
\begin{equation}
\label{eq:Jacobian_grad_t}
J\left(r,z\right)=\left(r\partial_{z}t+2z\partial_{r}t\right)\partial_{z}t-\left(2z\partial_{z}t-r\partial_{r}t\right)\partial_{r}t=r\left(\nabla t\right)^{2},
\end{equation}
where the nabla operator is understood to work in the \emph{Cartesian} coordinates $r,z$.
We now plug Eq.~(\ref{eq:Jacobian_grad_t}) into Eq.~(\ref{eq:action_rz_plane_with_Jacobian}), and recalling that $\tbar = 1-t$, we arrive at
\begin{equation}
\label{eq:action_in_hodograph_before_integrating_by_parts}
S=\frac{1}{2}\int_{-\infty}^{\infty}dz\int_{0}^{\infty}dr\,r^{5}\left(\nabla\tbar\right)^{2}.
\end{equation}
By virtue of Eq.~(\ref{eq:Laplace_like_t_z_r}), the integrand in Eq.~(\ref{eq:action_in_hodograph_before_integrating_by_parts}) can be written in the form
\begin{equation}
\label{eq:integrand_by_parts}
r^{5}\left(\nabla\tbar\right)^{2}=\nabla\cdot\left(r^{5}\tbar\nabla\tbar-r^{4}\tbar^{2}\hat{r}\right)+4r^{3}\tbar^{2},
\end{equation}
where $\hat{r}$ is a unit vector in the $r$ direction.
We plug Eq.~(\ref{eq:integrand_by_parts}) into~(\ref{eq:action_in_hodograph_before_integrating_by_parts}) and use Gauss' theorem in order to obtain
\begin{equation}
\label{eq:S_rzplane_with_boundary_terms}
S=\frac{1}{2}\oint\left(r^{5}\tbar\nabla\tbar-r^{4}\tbar^{2}\hat{r}\right)\cdot\vec{dn}+2\int_{-\infty}^{\infty}dz\int_{0}^{\infty}dr\,r^{3}\tbar^{2}
\end{equation}
where the first integral in Eq.~(\ref{eq:S_rzplane_with_boundary_terms}) is over the boundary of the integration domain in the $rz$ plane, with $\vec{dn}$ pointing outwards.
We first consider the boundary terms in Eq.~(\ref{eq:S_rzplane_with_boundary_terms}).
The boundary term on the $z$ axis,
\begin{equation}
\frac{1}{2}\left[\int_{-\infty}^{\infty}\left(r^{4}\tbar^{2}-r^{5}\tbar\partial_{r}\tbar\right)dz\right]_{r=0},
 \end{equation}
vanishes, because $t(r,z)$ has regular (nonsingular) behavior at $r \to 0$.
The boundary terms at infinity,
\begin{equation}
\frac{1}{2}\left[\int_{-\infty}^{\infty}\left(r^{5}\tbar\partial_{r}\tbar-r^{4}\tbar^{2}\right)dz\right]_{r=\infty}
 \end{equation}
 and
 \begin{equation}
\frac{1}{2}\left[\int_{0}^{\infty}\left(r^{5}\tbar\partial_{z}\tbar\right)dr\right]_{z=-\infty}^{z=\infty},
 \end{equation}
vanish because of the asymptotic decay~(\ref{eq:tbar_asymptotics_large_rz}) of $\tbar$. The only nonvanishing boundary terms are therefore those on the interval $r_1 \le r \le r_2$, on both sides of the $z$ axis. These terms can be written in the form:
\begin{equation}
\label{eq:boundary_terms_on_dirichlet}
-\frac{1}{2}\int_{r_{1}}^{r_{2}}\left[r^{5}\tbar\partial_{z}\tbar\left(r,z\right)\right]_{z=0^{-}}^{z=0^{+}}dr\simeq-\frac{\bar{r}^{5}}{2}\int_{r_{1}}^{r_{2}}\left[\partial_{z}\tbar\left(r,z\right)\right]_{z=0^{-}}^{z=0^{+}}dr,
\end{equation}
where we plugged in $\tbar=1$ on the boundary due to boundary condition~(\ref{eq:t_Dirichlet_condition}), and used the assumption $\Delta \ll \bar{r}$ in order to approximate $r \simeq \bar{r}$ in the integration domain. Therefore, the action is altogether given by:
\begin{equation}
\label{eq:S_rz_plane_after_integration_by_parts}
S\simeq-\frac{\bar{r}^{5}}{2}\int_{r_{1}}^{r_{2}}\left[\partial_{z}\tbar\left(r,z\right)\right]_{z=0^{-}}^{z=0^{+}}dr+2\int_{-\infty}^{\infty}dz\int_{0}^{\infty}dr\,\,r^{3}\tbar^{2}\left(r,z\right).
\end{equation}

We can find a convenient expression for the integrand which appears in the first integral in Eq.~(\ref{eq:S_rz_plane_after_integration_by_parts}), by using the fact that the difference between the electric fields on opposite sides of a charged surface is equal to the surface charge density multiplied by $-4\pi$. For our 3-dimensional electrostatic problem, the form of the charge density is given by Eq.~(\ref{eq:wdef}), so
\begin{equation}
\label{eq:dz_dt_on_the_boundary}
\left.\partial_{z}\tbar\left(r,z\right)\right|_{z=0^{-}}^{z=0^{+}}=\frac{\left.\partial_{z}\Psi\left(r,z,\phi\right)\right|_{z=0^{-}}^{z=0^{+}}}{r\cos\phi}=-\frac{4\pi g\left(r\right)}{r}\simeq-\frac{4\pi g\left(r\right)}{\bar{r}},\qquad r_{1}\le r\le r_{2} .
\end{equation}
We now plug Eq.~(\ref{eq:dz_dt_on_the_boundary}) into the first term in Eq.~(\ref{eq:S_rz_plane_after_integration_by_parts}), to obtain:
\begin{equation}
\label{eq:action_boundary_term_result}
-\frac{\bar{r}^{5}}{2}\int_{r_{1}}^{r_{2}}\left[\partial_{z}\tbar\left(r,z\right)\right]_{z=0^{-}}^{z=0^{+}}dr\simeq2\pi\bar{r}^{4}\int_{r_{1}}^{r_{2}}dr\,g\left(r\right)=2\pi\bar{r}^{4}I_{0},
\end{equation}
where we used Eq.~(\ref{eq:I0def}).
In the leading order, Eqs.~(\ref{eq:rbar_Delta_L}),~(\ref{eq:H_L_rbar}) and~(\ref{eq:action_boundary_term_result}) give the action of the uniform solution~(\ref{eq:action_uniform_solution}). The  corrections to this result are due to the subleading term in Eq.~(\ref{eq:H_L_rbar}), and due to the double integral in Eq.~(\ref{eq:S_rz_plane_after_integration_by_parts}), to which we now turn our attention.

First of all, we notice that the main contribution to the double integral in Eq.~(\ref{eq:S_rz_plane_after_integration_by_parts}) is from the region where Eq.~(\ref{eq:t_r_z_far_regime}) is valid
\footnote{The contribution to the double integral in Eq.~(\ref{eq:S_rz_plane_after_integration_by_parts}) from the region
$\sqrt{\left(r-\bar{r}\right)^{2}+z^{2}}=O(\Delta)$
can be estimated by replacing $r$ by $\bar{r}$ and $\tbar$ by $1$ in the integrand, and restricting the integration domain to a circle around $r=\bar{r},z=0$ whose radius is of order $\Delta$. The result is of order $\bar{r}^{3}\Delta^{2}$, which,  in the thin annulus limit $\Delta \ll \bar{r}$, is much smaller than the contribution~(\ref{eq:action_bulk_term_result}) which comes from the region $\sqrt{\left(r-\bar{r}\right)^{2}+z^{2}}\gg\Delta$.}.
 Plugging Eq.~(\ref{eq:t_r_z_far_regime}) into the double integral in Eq.~(\ref{eq:S_rz_plane_after_integration_by_parts}), which we denote $S_{\text{bulk}}$, yields
\begin{eqnarray}
\!\!\!\! S_{\text{bulk}}&=&2\int_{-\infty}^{\infty}dz\int_{0}^{\infty}dr\,\,r^{3}\tbar^{2}\left(r,z\right)\simeq2I_{0}^{2}\bar{r}^{2}\int_{-\infty}^{\infty}dz\int_{0}^{\infty}dr\,\,r\left[\mathcal{F}\left(r,\bar{r},z\right)\right]^{2}  \nonumber\\
&=&8I_{0}^{2}\bar{r}^{2}\int_{-\infty}^{\infty}d\tilde{z}\int_{0}^{\infty}d\tilde{r}\,\frac{\left\{ \left(\tilde{r}^{2}+1+\tilde{z}^{2}\right)\EllipticK\left[-\frac{4\tilde{r}}{\left(\tilde{r}-1\right)^{2}+\tilde{z}^{2}}\right]-\left[\left(\tilde{r}-1\right)^{2}+\tilde{z}^{2}\right]\EllipticE\left[-\frac{4\tilde{r}}{\left(\tilde{r}-1\right)^{2}+\tilde{z}^{2}}\right]\right\} }{\tilde{r}\left[\left(\tilde{r}-1\right)^{2}+\tilde{z}^{2}\right]}^{2},
\end{eqnarray}
where we rescaled the integration variables by defining $\tilde{r} = r / \bar{r}$ and $\tilde{z} = z / \bar{r}$. After an additional change of variables,
\begin{eqnarray}
u&=&\sqrt{\tilde{z}^{2}+\left(\tilde{r}-1\right)^{2}} ,\\
v&=&\frac{4\tilde{r}}{\tilde{z}^{2}+\left(\tilde{r}-1\right)^{2}},
\end{eqnarray}
$S_{\text{bulk}}$ takes the form
\begin{equation}
\label{eq:S_integral_uv}
S_{\text{bulk}}=16I_{0}^{2}\bar{r}^{2}\int_{0}^{\infty}dv\int_{u_{\min}\left(v\right)}^{u_{\max}\left(v\right)}du\,\frac{u^{3}\left[\left(\frac{v}{2}+1\right)\EllipticK\left(-v\right)-\EllipticE\left(-v\right)\right]}{v\sqrt{u^{2}-\left(\frac{u^{2}v}{4}-1\right)^{2}}}^{2},
\end{equation}
where
\begin{equation}
u_{\min}\left(v\right)\equiv\frac{2\left(-1+\sqrt{1+v}\right)}{v},\quad u_{\max}\left(v\right)\equiv\frac{2\left(1+\sqrt{1+v}\right)}{v}.
\end{equation}
We can now evaluate the integrals in Eq.~(\ref{eq:S_integral_uv}) and obtain
\begin{equation}
\label{eq:action_bulk_term_result}
S_{\text{bulk}}=32\pi I_{0}^{2}\bar{r}^{2}\int_{0}^{\infty}dv\:\frac{\left(2+v\right)\left[\left(2+v\right)\EllipticK\left(-v\right)-2\EllipticE\left(-v\right)\right]^{2}}{v^{4}}=\frac{32\pi I_{0}^{2}\bar{r}^{2}}{3}.
\end{equation}
The action $S$ is given by the the sum of Eqs.~(\ref{eq:action_boundary_term_result}) and~(\ref{eq:action_bulk_term_result}):
\begin{equation}
\label{eq:S_L_rbar}
S\simeq2\pi\bar{r}^{4}I_{0}+\frac{32\pi I_{0}^{2}\bar{r}^{2}}{3}.
\end{equation}
We can now use Eqs.~(\ref{eq:rbar_Delta_L}),~(\ref{eq:H_L_rbar}) and~(\ref{eq:S_L_rbar}) to express the action in terms of $H$ and $\ell$, yielding Eq.~(\ref{eq:S_H_L_positive_tail}).

\section{The optimal height profile for $\lambda H \to -\infty$}
\subsubsection{The intermediate stage of the dynamics}
\label{Appendix:optimal_profile_positive_tail_without_end}

\renewcommand{\theequation}{E\arabic{equation}}
\setcounter{equation}{0}

In the intermediate regime, $\Delta\ll\left|z\right|$, $\left|r-\bar{r}\right|\ll\bar{r}$,
Eq.~(\ref{eq:t_r_z_far_regime}) takes the form:
\begin{equation}
\label{eq:t_r_z_intermediate_region}
t\left(r,z\right)\simeq1-\frac{I_{0}}{\bar{r}}\ln\left\{ \frac{64\bar{r}^{2}}{e^{4}\left[\left(r-\bar{r}\right)^{2}+z^{2}\right]}\right\}.
\end{equation}
Plugging Eq.~(\ref{eq:t_r_z_intermediate_region}) into Eqs.~(\ref{eq:OFM_dx_dz}) and~(\ref{eq:OFM_dx_dr}) and integrating the latter equations yields
\begin{equation}
\label{eq:x_r_z_intermediate_region}
x\left(r,z\right)\simeq x_{0}-2I_{0}\,\text{arctan}\left(\frac{z}{r}\right).
\end{equation}
Eqs.~(\ref{eq:t_r_z_intermediate_region}) and~(\ref{eq:x_r_z_intermediate_region}) can be inverted to obtain
\begin{eqnarray}
\label{eq:r_x_t_intermediate_region}
r\left(x,t\right)&=&\,\sqrt{H}\left(1-\frac{2\ell}{\pi\sqrt{H}}\right)+8\sqrt{H}\exp\left[\frac{\pi\sqrt{H}}{\ell}\left(t-1\right)-2t\right]\cos\left(\frac{\pi x}{\ell}\right), \\
\label{eq:z_x_t_intermediate_region}
z\left(x,t\right)&=&-8\sqrt{H}\exp\left[\frac{\pi\sqrt{H}}{\ell}\left(t-1\right)-2t\right]\sin\left(\frac{\pi x}{\ell}\right),
\end{eqnarray}
where we used Eqs.~(\ref{eq:rbar_Delta_L}) and~(\ref{eq:H_L_rbar}) in order to express the result in terms of $\ell$ and $H$, and we chose $x_0=0$ in order to ensure that the maximal height interface is attained at $x=0$.
We now integrate Eq.~(\ref{eq:z_x_t_intermediate_region}) with respect to $x$,
\begin{equation}
\label{eq:h_x_t_with_Ct}
h\left(x,t\right)\simeq C\left(t\right)+\frac{16\ell\sqrt{H}}{\pi}\exp\left[\frac{\pi\sqrt{H}}{\ell}\left(t-1\right)-2t\right]\cos\left(\frac{\pi x}{\ell}\right).
\end{equation}
Plugging Eqs.~(\ref{eq:r_x_t_intermediate_region}) and~(\ref{eq:h_x_t_with_Ct}) into~(\ref{eq:OFM_h}), and taking into account the flat initial condition, we find the function $C(t)$
\begin{equation}
C\left(t\right)=\bar{r}^{2}t\simeq H\left(1-\frac{4\ell}{\pi\sqrt{H}}\right)t,
\end{equation}
altogether yielding the optimal height profile~(\ref{eq:optimal_profile_positive_tail_without_end}).

\subsubsection{The optimal height profile at the end of the dynamics}
\label{Appendix:final_optimal_profile_positive_tail}

The hodograph transformation solves the OFM equations only inside an interval of $x$ where $\rho>0$ \citep{MKV}. This interval is $\left|x\right| \le \Hopfell\left(t\right)$, where $\Hopfell\left(t<t_c\right) = \ell$, and $t_c$ is a critical time found below.  In the so called Hopf region (defined by $t > t_c$ and $\ell \ge \left|x\right|>\Hopfell\left(t\right)$), $\rho$ vanishes
(to remind the reader, we neglect the diffusion terms here) and therefore Eq.~(\ref{eq:OFM_V_positive}) reduces to the Hopf equation $\partial_{t}V+V\partial_{x}V = 0$. The general solution to the Hopf equation can be written in an implicit form \citep{LL}:
\begin{equation}
\label{eq:F_xVt}
F\left(V\right)=x-Vt.
\end{equation}
The function $F$ in Eq.~(\ref{eq:F_xVt}) is found from matching the hodograph and Hopf solutions at $\left|x\right|=\Hopfell\left(t\right)$, $t_{c}\le t\le1$, where $V$ must be continuous. This boundary corresponds to the $z$ axis of the hodograph plane. Using Eqs.~(\ref{eq:rbar_Delta_L}) and~(\ref{eq:H_L_rbar}), Eq.~(\ref{eq:t_on_z_axis}) gives us a convenient expression for $t$ on the $z$ axis in terms of $H$ and $\ell$:
\begin{equation}
\label{eq:t_on_z_axis_L_rbar}
t\left(r=0,z\right)\simeq1-\frac{\ell H}{2\left(H+z^{2}\right)^{3/2}}.
\end{equation}
By plugging $z=0$ into Eq.~(\ref{eq:t_on_z_axis_L_rbar}), we obtain the critical time $t_c$ at which $\rho$ vanishes at $x=\pm \ell$: $t_{c}=t\left(r=0,z=0\right)\simeq1-\ell/\left(2\sqrt{H}\right)$.
In order to find $x(r,z)$ on the $z$ axis,
we plug Eq.~(\ref{eq:t_on_z_axis_L_rbar}) into Eq.~(\ref{eq:dx_dz_on_z_axis}) and then integrate the latter equation with respect to $z$, together with the boundary condition $x\left(r=0,z\to\pm\infty\right)=0$,
 and we obtain
 \begin{equation}
 \label{eq:x_on_z_axis_L_rbar}
x\left(r=0,z\right)\simeq\frac{\ell z^{3}}{\left(H+z^{2}\right)^{3/2}}-\ell\,\text{sgn}\left(z\right).
 \end{equation}
The function $F$ in Eq.~(\ref{eq:F_xVt}) is now found by plugging Eqs.~(\ref{eq:t_on_z_axis_L_rbar}) and~(\ref{eq:x_on_z_axis_L_rbar}) into Eq.~(\ref{eq:F_xVt}), and using the connection $z=V/2$, yielding an implicit solution for $V(x,t)$ in the Hopf region
\begin{equation}
\label{V_xt_Hopf_particular_implicit}
x-Vt=\frac{\ell V}{\sqrt{4H^{2}+V^{2}}}-\ell\,\text{sgn}\left(V\right)-V,\qquad t_{c}\le t\le1,\;\Hopfell\left(t\right)\le\left|x\right|\le \ell.
\end{equation}
Eq.~(\ref{V_xt_Hopf_particular_implicit}) can, in principle, be solved for $V$, and then the solution should be matched with the solution in the hodograph region in order to obtain the optimal height profile $h(x,t)$ in the Hopf region. This calculation is quite difficult for arbitrary times $t_c \le t \le 1$, because the hodograph solution for $h(x,t)$ is known only in implicit form. However, finding the optimal profile at the \emph{end} of the dynamics is easy: plugging $t=1$ into Eq.~(\ref{V_xt_Hopf_particular_implicit}) yields
\begin{equation}
\label{V_x_final_explicit}
V\left(x,t=1\right)=\frac{2\sqrt{H}\left[x-\ell\,\text{sgn}\left(x\right)\right]}{\sqrt{x\left[2\ell\,\text{sgn}\left(x\right)-x\right]}}.
\end{equation}
It follows from the boundary condition~(\ref{eq:delta_initial_cond_rho}) that $\Hopfell\left(t=1\right)=0$. Eq.~(\ref{V_x_final_explicit}) is therefore valid for all $x$, and it can be immediately integrated with respect to $x$, with the boundary condition $h(0,1)=H$, to obtain the optimal profile at $t=1$~(\ref{eq:final_optimal_profile_positive_tail}).

\section{The weakly supercritical inviscid limit}

\renewcommand{\theequation}{F\arabic{equation}}
\setcounter{equation}{0}

For the purposes of this Appendix we change the notation from $z$ and $r$ to $X$ and $Y$, so that Eq.~(\ref{eq:Laplace_like_t_z_r}) becomes
\begin{equation}
\label{eq:Laplace_like_t_X_Y}
\partial_{X}^{2}t+\partial_{Y}^{2}t+\frac{3\partial_{Y}t}{Y}=0.
\end{equation}
Correspondingly, in Eq.~(\ref{eq:t_Dirichlet_condition}), $r_{1,2}$ are replaced by $Y_{1,2}$ . We assume here $Y_1 \ll Y_2$.

\subsubsection{Matched asymptotics}
\label{appendix:small_hole_matched_asymptotics}
Sufficiently far from the hole, $X^2 + Y^2 \gg Y_1^2$, an approximate solution to Eq.~(\ref{eq:Laplace_like_t_X_Y}) is found simply by setting $Y_1 = 0$. The resulting
problem was solved, in the context of the statistics of integrated current throuh a lattice site of the SSEP, in Ref. \cite{MS2014}. The solution
(which plays the role of the outer solution to our problem) is
\begin{equation}
\label{outersolution}
t_{\text{outer}}\left(s\right)=\frac{2}{\pi}\left(\frac{s}{1+s^{2}}
+\text{arctan}\,s\right), \qquad X^2 + Y^2 \gg Y_1^2,
\end{equation}
where
$$
s = \left[\frac{X^{2}+Y^{2}-Y_{2}^{2}+\sqrt{\left(X^{2}+Y^{2}-Y_{2}^{2}\right)^{2}+4X^{2}Y_{2}^{2}}}{2Y_{2}^{2}}\right]^{1/2}.
$$
Far from the disk edge, $X^2 + Y^2 \ll Y_2^2$, we can set $Y_2 = \infty$ and employ the elliptic coordinates
\begin{eqnarray}
X&=&Y_1\se\re,\\
Y&=&Y_1\sqrt{\left(1+\se^{2}\right)\left(1-\re^{2}\right)},
\end{eqnarray}
where $\se \ge 0$ and $\left|\re\right|\le1$.
In these coordinates, Eq.~(\ref{eq:Laplace_like_t_X_Y}) becomes
\begin{equation}
\label{eq:Laplace_like_elliptic}
\partial_{\se}\left[\left(1+\se^{2}\right)^{2}\partial_{\se}t\right]+\frac{1+\se^{2}}{1-\re^{2}}\partial_{\re}\left[\left(1-\re^{2}\right)^{2}\partial_{\re}t\right]=0.
\end{equation}
The condition~(\ref{eq:t_Dirichlet_condition}) becomes, for $Y_2 = \infty$,
\begin{equation}
\label{eq:Laplace_like_elliptic_BC}
t\left(\se,\re=0\right)=0.
\end{equation}
Now we demand that the solution to the problem~(\ref{eq:Laplace_like_elliptic}) and~(\ref{eq:Laplace_like_elliptic_BC}) match with the outer solution (\ref{outersolution}) in the joint region of their validity $Y_1^2 \ll X^2 + Y^2 \ll Y_2^2$. We obtain
\begin{equation}
t_{\text{inner}}=\frac{8Y_{1}\left|\re\right|}{\pi^{2}Y_{2}}\left(\se \, \text{arctan}\,\se+\frac{\se^{2}+\frac{2}{3}}{\se^{2}+1}\right), \qquad X^2 + Y^2 \ll Y_2^2.
\end{equation}
The two solutions indeed match: in the joint region they are $t_{\text{inner}}\simeq t_{\text{outer}}\simeq4\left|X\right|/\left(\pi Y_{2}\right)$.

\subsubsection{Evaluating $L/\sqrt{t}$, $H$ and $S$ as functions of $Y_1$ and $Y_2$}
\label{appendix:small_hole_ell_H_S}

We choose an arbitrary $Y_{1}\ll Y_{m}\ll\sqrt{Y_{1}Y_{2}}$, and using Eq.~(\ref{eq:L_as_a_hodograph_integral}),  we obtain
\begin{eqnarray}
\label{eq:ell_Y1_Y2}
\ell&=&2\int_{0}^{\infty}\!\!\tbar\left(Y=0,X\right)dX\simeq2\left[\int_{0}^{Y_{m}}\!\tbar_{\text{inner}}\left(Y=0,X\right)dX+\int_{Y_{m}}^{\infty}\!\tbar_{\text{outer}}\left(Y=0,X\right)dX\right]\nonumber\\
&=&\frac{2}{\pi}\left[2Y_{2}+\pi Y_{m}-2Y_{m}\text{arctan}\left(\frac{Y_{2}}{Y_{m}}\right)-\frac{\left(4Y_{1}^{2}+12Y_{m}^{2}\right)\text{arctan}\left(\frac{Y_{m}}{Y_{1}}\right)+12Y_{1}Y_{m}}{3\pi Y_{2}}\right],
\end{eqnarray}
which yields Eq.~(\ref{eq:ell_r1_r2_main_text}) of the main text (where $r_i = Y_i$).
In the leading order, Eq.~(\ref{eq:ell_r1_r2_main_text}) is independent of $Y_m$, as expected.
Using Eq.~(\ref{eq:H_integral_rz_plane}), we find
\begin{eqnarray}
\label{eq:H_Y2}
H&=&Y_{2}^{2}+2\int_{Y_{2}}^{\infty}Y\tbar\left(X=0,Y\right)dY\simeq Y_{2}^{2}+2\int_{Y_{2}}^{\infty}Y\tbar_{\text{outer}}\left(X=0,Y\right)dY\nonumber\\
&=&Y_{2}^{2}+2Y_{2}^{2}\int_{1}^{\infty}\eta\left\{ 1-\frac{2}{\pi}\left[\frac{\sqrt{\eta^{2}-1}}{\eta^{2}}+\text{arctan}\left(\sqrt{\eta^{2}-1}\right)\right]\right\} d\eta.
\end{eqnarray}
After evaluating the integral in Eq.~(\ref{eq:H_Y2}), we obtain Eq.~(\ref{eq:H_r2_main_text}) of the main text (where $r_2 = Y_2$).

For $Y_1 = 0$, the action can be found by evaluating Eq.~(\ref{eq:S_rzplane_with_boundary_terms}) on $\tbar_{\text{outer}}$. The result is
\begin{equation}
S_{\text{boundary}}\left(Y_{1}=0,Y_{2}\right)=S_{\text{bulk}}\left(Y_{1}=0,Y_{2}\right)=\frac{32Y_{2}^{5}}{15\pi},
\end{equation}
yielding Eq.~(\ref{eq:S_Y2_zero_order}) of the main text.
In order to evaluate the $Y_1\!>\!0$ corrections to Eq.~(\ref{eq:S_Y2_zero_order}), the domains of integration in the calculations of $S_{\text{boundary}}$ and of $S_{\text{bulk}}$ are divided into inner and outer regions in a manner similar to Eq.~(\ref{eq:ell_Y1_Y2}). However, one must also account for the leading correction to $t_\text{outer}$. This correction must match the subleading term of the asymptotic expansion of $t_\text{inner}$ in the joint region,
\begin{equation}
t_{\text{inner}}\simeq\frac{4\left|X\right|}{\pi Y_{2}}\left[1+\frac{4}{15\pi}\left(\frac{Y_{1}^{2}}{X^{2}+Y^{2}}\right)^{5/2}\right],\qquad Y_{1}\ll X^{2}+Y^{2}\ll Y_{2}^{2}.
\end{equation}
We deduce from here that the leading order correction to the outer solution should be of the fifth order in $Y_1$:
\begin{equation}
\label{t_outer_correction_scaling}
t_{\text{outer}}=t_{\text{outer}}^{\left(0\right)}\left(\frac{X}{Y_{2}},\frac{Y}{Y_{2}}\right)+\left(\frac{Y_{1}}{Y_{2}}\right)^{5}t_{\text{outer}}^{\left(1\right)}\left(\frac{X}{Y_{2}},\frac{Y}{Y_{2}}\right).
\end{equation}
Now Eqs.~(\ref{eq:S_rzplane_with_boundary_terms}) and~(\ref{t_outer_correction_scaling}) imply that the $Y_1\!>\!0$ corrections to the action are of the form given in Eq.~(\ref{eq:S_Y1_scaling}).

In order to make further analytical progress here, $t_{\text{outer}}^{\left(1\right)}$ must be expanded over the eigenmodes of Eq.~(\ref{eq:Laplace_like_t_X_Y}).
Instead of performing this cumbersome calculation, we solved Eq.~(\ref{eq:Laplace_like_t_X_Y}) numerically, using an artificial relaxation method, and then computed the action on the numerical solutions using Eq.~(\ref{eq:S_rzplane_with_boundary_terms}). Our numerical results are plotted in Fig. \ref{fig:smallHole}.

\end{appendices}

\end{document}